%% file: main.tex
\documentclass[aps,pra,twocolumn,superscriptaddress,showpacs]{revtex4-2}

\usepackage{graphicx}
\usepackage{amsmath}
\usepackage{amssymb}
\usepackage{amsfonts}
\usepackage{hyperref}
\usepackage{xcolor}
\usepackage{wrapfig}  
\usepackage{lipsum}  
\usepackage{bm}
\usepackage{comment}
\usepackage{soul}
\setstcolor{red}
\usepackage{appendix}
\usepackage{booktabs} 
\newcommand{\rl}[1]{{\color{cyan}{#1}}}
\newcommand{\as}[1]{{#1}}
\newcommand{\ta}[1]{{{#1}}}

\newcommand{\q}[1]{\vert #1 \rangle}

\newcommand\pza{\varphi_{\text{zpf},a}}

\newcommand\pzb{\varphi_{\text{zpf},b}}
\newcommand{\bigdot}[1]
{\overset{\text{\LARGE$\cdot$}}{#1}}

\begin{document}


\title{A cat qubit stabilization scheme using a voltage biased Josephson junction}




\author{T.Aissaoui}
\affiliation{Alice \& Bob, 53 Bd du Général Martial Valin, 75015 Paris, France}
\affiliation{Laboratoire de Physique de l’École Normale Supérieure, Inria, ENS,
Mines ParisTech, Université PSL, Sorbonne Université, Paris, France}

\author{A.Murani}
\affiliation{Alice \& Bob, 53 Bd du Général Martial Valin, 75015 Paris, France}

\author{R.Lescanne}
\affiliation{Alice \& Bob, 53 Bd du Général Martial Valin, 75015 Paris, France}

\author{A.Sarlette}
\affiliation{Laboratoire de Physique de l’École Normale Supérieure, Inria, ENS,
Mines ParisTech, Université PSL, Sorbonne Université, Paris, France}


\begin{abstract}
DC-voltage-biased Josephson junctions have been recently employed in superconducting circuits for Hamiltonian engineering, demonstrating microwave amplification, single photon sources and entangled photon generation. Compared to more conventional approaches based on parametric pumps, this solution typically enables larger interaction strengths. In the context of quantum information, a two-to-one photon interaction can stabilize cat qubits, where bit-flip errors are exponentially suppressed, promising significant resource savings for quantum error correction. This work investigates how the DC bias approach to Hamiltonian engineering can benefit cat qubits. We find a simple circuit design that is predicted to showcase a two-to-one photon exchange rate larger than that of the parametric pump-based implementation while dynamically averaging typically resonant parasitic terms such as Kerr and cross Kerr. In addition to addressing qubit stabilization, we propose to use injection locking with a cat-qubit adapted frequency filter to prevent long-term drifts of the cat qubit angle associated to DC voltage noise. The whole scheme is simulated without rotating-wave approximations, highlighting for the first time the amplitude of related oscillatory effects in cat-qubit stabilization schemes. This study lays the groundwork for the experimental realization of such a circuit. 
\end{abstract}

\maketitle


\input{sections/1_introduction}

\input{sections/2_model}
\input{sections/3_noise}

\input{sections/4_discussion}
\input{sections/5_conclusion}

\begin{acknowledgments}
    The authors wish to thank Ambroise Peugeot, Lukas Danner, Florian Höhe, Ciprian Padurariu, Joachim Ankerhold and Björn Kubala for the insightful discussions.
    
    AS acknowledges financial support from the ANR, project ANR-JCJC HAMROQS and Plan France 2030 ANR-22-PETQ-0006.
\end{acknowledgments}

\bibliography{sections/bibliography}
\bibliographystyle{apsrev4-2}

\input{sections/appendix}

\end{document}

%% file: sections/1_introduction.tex
\section{Introduction}\label{sec:intro}

Superconducting circuits \cite{Blais2021} provide an ideal platform to develop bosonic codes \cite{Joshi2021}, which are designed to reduce the hardware requirements for quantum error correction by moving beyond traditional two-level systems \cite{Krinner2022, Google2023}. One prominent example of such codes is the cat qubit \cite{Mirrahimi2014, Puri2017}, encoded in two coherent states of a harmonic oscillator with identical mean photon number but opposite amplitudes. This mesoscopic encoding, leveraging distant states in phase space, ensures exponential protection against bit-flip errors at a linear cost in phase-flip errors as the mean photon number increases \cite{Lescanne2020}. To attain this protection against typical loss channels, the two coherent basis states are stabilized by a two-photon drive-and-dissipation mechanism \cite{Leghtas2015}. \as{Crucially, this process does not induce phase-flip errors, as it does not distinguish between the two basis states. 
The stabilization rate then sets the timescale for protecting quantum information against losses and against gate-induced imperfections \cite{Gautier2023,Guillaud2019}. }
Therefore, to perform fast and high fidelity gates, a strong stabilization is required. 

At the heart of cat qubit stabilization schemes lies a two-to-one photon exchange Hamiltonian between the high Q mode hosting the cat qubit, called the memory, and a low Q mode called the buffer. In superconducting circuits, engineering such Hamiltonian relies on the non-linearity of one or several Josephson junctions, tuned with external RF or DC biases. A good design results in a strong exchange Hamiltonian yielding large stabilization rate and hence high fidelity operations, while mitigating the spurious terms that induce bit-flip or phase-flip errors.
Several strategies have already been explored to implement this three-wave mixing interaction. One approach \cite{Leghtas2015, Touzard2018} uses the four wave-mixing properties of a single Josephson junction and a parametric pump at the frequency that makes the two-to-one photon exchange resonant. However, a strong cross-Kerr interaction between the memory and buffer mode limits the achievable bit-flip times. A second strategy \cite{Lescanne2020, Berdou, Reglade2024, putterman2024} eliminates such parasitic terms by symmetry. This method employs a pair of Josephson junctions embedded in an asymmetrically threaded SQUID (ATS), which is flux-pumped to satisfy the resonance condition. While this device demonstrates exponential bit-flip suppression, the dynamics of the system under strong pump, required for strong interaction, are not yet well understood. Additionally, operating this device is more complex due to the presence of two flux loops and two DC and RF biases. A third approach \cite{Marquet2023, Marquet2024} involves using a three-wave mixing device that essentially consists of a current biased Josephson junction. This device exhibits strong interaction strength, however the mode frequencies themselves must be tuned for the target interaction to be resonant, which reduces flexibility in frequency arrangement. The use of DC voltage bias as an additional knob in such designs was also mentioned in \cite{rojkov2024stabilization}. 

Recent advances in Josephson photonics \cite{Hofheinz2011, Cassidy2017, Jebari2018, Rolland2019, Peugeot2021, Menard2022, Albert2024} have activated parametric interactions by only applying a DC voltage across a Josephson junction. Instead of modulating the phase across the junction with an AC pump, the DC voltage makes it wind around at a frequency proportional to the voltage, thereby inducing periodic oscillation of the Josephson energy between $E_J$ and $-E_J$ (the so-called AC Josephson effect).
This method not only provides strong modulation but also averages out most spurious terms. In this work, we explore the use of this approach to engineer the two-to-one photon exchange Hamiltonian required for stabilizing cat-qubits in a novel circuit design which we refer to as the \textit{DC cat qubit}. We establish the ability for such circuit to stabilize cat qubits and analyze its expected advantage compared to previous designs. We also characterize the impact of voltage noise, that we propose to mitigate by employing injection locking \cite{Balanov, Pikovsky2001, Cassidy2017, Bengtsson2018, Markovic2019, Danner2021, LakshmiBhai2023}, in an adapted version to remain compatible with cat-qubit protection. We confirm the device performance by simulation, for the first time (to our knowledge) including the full time dependence without rotating wave approximations (Appendix \ref{appendix:quantum-simulation}). This highlights the importance of oscillatory effects and of dissipation filters in such devices.



%% file: sections/2_model.tex
\section{Ideal system}

\begin{figure}[htb]
\includegraphics[width=\linewidth]{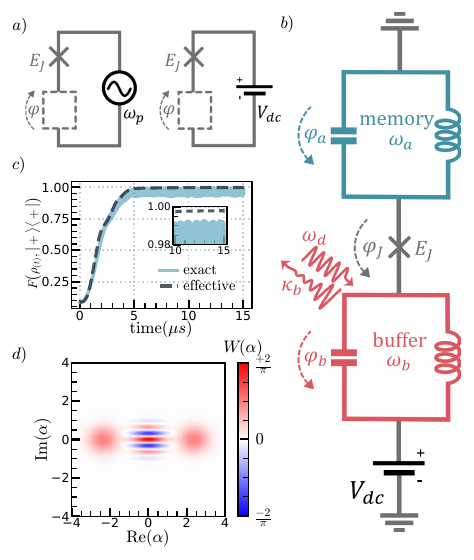}
\caption{\textbf{(a)} Schematic comparison of the Josephson parametric pumping approach (left) and the DC bias approach (right). The open box represents the rest of the circuit that should have a finite impedance at 0 frequency in the DC case.
\textbf{(b)} Lumped-element model of the proposed DC cat qubit circuit. A Josephson junction with energy $E_J$ is put in series with the memory (blue) and buffer (red) resonators. A voltage source closes the circuit and provides the DC voltage bias $V_{dc}$ that activates the two-to-one photon interaction between the memory and buffer resonators when $ 2 e V_{dc}/\hbar = 2\omega_a - \omega_b$. To stabilize cat qubits, and similarly to previous schemes, the buffer resonator is coupled to a dissipative environment and driven by a microwave source (arrows). \textbf{(c)} Time evolution of the memory state fidelity to an even cat state with 5.5 photons, when turning on the stabilization dynamics starting from vacuum. The agreement between the effective dynamics Hamiltonian (with Taylor development up to order 5 in $\varphi_{\text{zpf}}$ and other approximations, see Appendix \ref{appendix:eff_Hamiltonian};black dashed) and the exact time-dependent one (Eq.~\eqref{eq:H_time} plus filtered buffer dissipation \ref{appendix:quantum-simulation}; blue curve) validates the approach. The blue curve thickness arises from fast oscillations dominated by the off-resonant displacement at frequency $\omega_{dc}$. The final infidelity (inset) is below 1\% and is attributed to parity loss occurring during the transient phase of the dynamics. \textbf{(d)} Wigner representation of the final state of the memory mode for the exact integration shown in (c).}
\label{fig:fig1}
\end{figure}

\paragraph*{{Cat qubit stabilization}} \message{\showthe\columnwidth} Stabilizing a cat qubit manifold spanned by the coherent states $\{|\alpha\rangle, |-\alpha\rangle\}$ requires engineering the following two-mode dynamics:
\begin{equation} \label{eq:H_cat}
    \begin{aligned}
        H_\mathrm{cat}/\hbar =& g_2 a^{\dag 2} b + g_2^* a^2 b^{\dag} + \epsilon_d b^{\dagger} + \epsilon_d^* b \; , \\
        L_b =& \sqrt{\kappa_b} b \; ,
    \end{aligned}
\end{equation}
where $a$ and $b$ represent the annihilation operators of the memory (high-Q) mode and buffer (low-Q)   mode respectively, $g_2$ denotes the rate of the two-to-one photon exchange dynamics, $\epsilon_d$ is the buffer drive. \as{The buffer dissipation at rate $\kappa_b$ is used to stabilize the cat qubit, while the memory mode is ideally assumed to have negligible dissipation.} One easily checks that this dynamics features an invariant space spanned by $\text{span}\{|\alpha\rangle, |-\alpha\rangle\}$, with $\alpha^2 = -\epsilon_d/g_2^*$ \cite{Mirrahimi2014}. When $g_2 \ll \kappa_b$, the two-to-one photon exchange, combined with the strong buffer dissipation, results in an effective two-photon dissipation on the memory with rate $\kappa_2 = 4|g_2|^2/\kappa_b$. The buffer drive inputs energy into the system and controls the amplitude of the cat qubit. Buffer dissipation and drive are commonly obtained by coupling the resonator to a dissipative load (50$\Omega$ line in most circuits) and a microwave source near its resonance frequency, respectively. In the following, we focus on engineering the two-to-one photon exchange Hamiltonian, that usually requires a parametric pump \cite{Leghtas2015, Touzard2018} with frequency 
\begin{equation}\label{eq:matching}
\omega_p = \omega_b - 2\omega_a
\end{equation}
to match the gap between the buffer frequency $\omega_b$ and twice the memory frequency $\omega_a$. As an alternative to this parametric pump, we propose using a DC-voltage biased Josephson junction \cite{Meister2015}.\\

\paragraph*{{RF pumping versus DC biasing}} 
To motivate our study, let's compare the basic dynamics of each approach. In the circuits depicted in Fig.~\ref{fig:fig1}a, the potential energy of the Josephson junction in the pumped and DC cases writes, respectively, 
\begin{align*}\label{eq:p_vs_dc}
    U_\mathrm{p}(\varphi) &= -E_J \cos(\epsilon_p \cos (\omega_p t ) + \varphi) \; , \\
    U_\mathrm{dc}(\varphi) &= -E_J \cos(\omega_{dc} t + \varphi) \; ,
\end{align*}
where $E_J$ is the Josephson energy and $\varphi$ is a general variable describing degrees of freedom in series with the junction. As the various modes coupled to the variable $\varphi$ will oscillate at their respective frequencies, the time-dependent signal is meant to select specific nonlinear effects --- in our application, two-to-one photon interaction --- through a frequency matching condition. In the parametric pump case, $\epsilon_p$ and $\omega_p$ represent respectively the pump amplitude and frequency. In the DC case, $\hbar \omega_{dc}/2 e = V_{dc}$ is the source voltage where $\hbar$ is the reduced Planck constant and $e$ the electron charge. 
Developing these equations, we understand how the parametric interaction occurs in both cases. In the parametric pump case, we have
\begin{equation} \label{eq:p_dev}
\begin{split}
    U_\mathrm{p}(\varphi) =& -E_J J_0(\epsilon_p)\cos( \varphi) \\
    & + 2 E_J J_1(\epsilon_p)\cos(\omega_p t)\sin(\varphi) \\
    & \ta{+ 2 E_J J_2(\epsilon_p)\cos(2\omega_p t)\cos(\varphi)} \\
    & + \cdots
\end{split}
\end{equation}
where \ta{$J_{i}$ are the $i$-th} Bessel functions of the first kind and where the dots denote terms oscillating at higher frequencies than $\omega_p$. In the DC case, we have
\begin{equation} \label{eq:dc_dev}
\begin{split}
U_\mathrm{dc}(\varphi) =& -E_J \cos(\omega_{dc} t )\cos( \varphi) \\
    & + E_J \sin(\omega_{dc} t)\sin(\varphi)\,.
\end{split}
\end{equation}
\as{The term performing two-to-one photon exchange would be the second one in each equation, while the remaining are spurious terms \cite{Leghtas2015}.
There are two key differences in this regard between equations \eqref{eq:p_dev} and \eqref{eq:dc_dev}.} First, the parametric pump case features many harmonics of $\omega_p$, which may induce spurious processes if we do not limit the pump amplitude, whereas the DC case contains only the frequency $\omega_{dc}$. 
Second, the two setups address characteristic frequencies inside $\cos(\varphi)$ differently. The parametric pump case features a residual stationary part (first line of Eq.\eqref{eq:p_dev}), activating non-rotating parts inside even powers of $\varphi$, like Kerr and cross-Kerr, \as{which are harmful to the dissipative cat stabilization \cite{Lescanne2020}} In the DC case, those terms are averaged out. Instead, the first line of Eq.\eqref{eq:dc_dev} would activate\as{ terms inside even powers of $\varphi$,} which specifically rotate at frequencies close to $\pm\omega_{dc}$. Such terms can involve more complicated processes, but can better be avoided by a proper selection of frequencies.\\


\paragraph*{{Circuit design}} For cat qubit stabilization, the voltage biased junction is put in series with two resonators as shown in Fig.~\ref{fig:fig1}b, such that $\varphi = \pza (a+ a^\dag) + \pzb (b+b^\dag)$ where $\varphi_{\text{zpf},a/b}$ are the zero point fluctuations of the phase across the memory and buffer resonators respectively. The time-dependent Hamiltonian of the system then writes 
\begin{equation}\label{eq:H_time}
\begin{split}
    H &= \hbar\omega_a a^{\dagger} a + \hbar \omega_bb^{\dagger} b + 2\text{Re}\left(\hbar{\epsilon_d} 
    e^{-i\omega_d t}\right)(b+b^{\dagger}) \\ - E_J &\cos \left[ \omega_{dc}t - \pza ( a+a^{\dagger}) - \pzb ( b+b^{\dagger} )  \right]\,.
\end{split}
\end{equation}
where the last term of the first line is the buffer drive at frequency $\omega_d = \omega_b$. 

To derive the more familiar two-to-one photon exchange Hamiltonian (see appendix \ref{appendix:eff_Hamiltonian}), we first move to the rotating frame of each resonator. Then, we perform a Taylor expansion of the trigonometric functions in $\varphi$ and, under the condition $\omega_{\text{dc}}, \omega_a, \omega_b, \omega_d \gg E_J \varphi_{\text{zpf},a/b}$, a Rotating Wave Approximation (RWA), averaging out rotating terms. Setting $\omega_{dc}$ at the frequency matching condition $\omega_{dc}=\omega_p$ from Eq.\eqref{eq:matching}, 
this yields as a leading term the two-to-one photon exchange of Eq.~\eqref{eq:H_cat} with $\hbar g_2 = \tilde{E}_J \varphi_a^2 \varphi_b /4i$
where $\tilde{E}_J = E_J \exp(-\pza^2/2-\pzb^2/2)$ \cite{Peugeot2021}. Refining the analysis, residual cross-Kerr and Kerr terms \as{can} arise from a quadratic effect on oscillatory linear displacements of the modes $a$ and $b$, \as{but not before order $(\varphi_{\text{zpf},a/b})^6$} (see Appendix \ref{appendix:Kerr}). Their amplitude is \as{thus several orders} of magnitude lower than $g_2$ thanks to the averaging effect of the DC voltage bias.\\

\paragraph*{{Numerical results}} To capture the contributions at all orders of averaging, we numerically simulate not only the usual RWA result, but also the time-dependent Hamiltonian Eq.~\eqref{eq:H_time} in presence of filtered buffer dissipation, and compare it to the basic cat qubit dynamics of Eq.~\eqref{eq:H_cat}. This also allows us to cover higher values of $\varphi_{\text{zpf},a/b}$, see Table \ref{table_1}.
The dissipation is focused in the vicinity of the buffer mode frequency (see Appendix \ref{appendix:quantum-simulation}), a constraint that is common to all cat stabilization approaches and is achieved with standard microwave filtering methods \cite{Chamberland2022}.
In Fig.~\ref{fig:fig1}, we show the preparation of an even cat state starting from vacuum when these dynamics are turned on. We find that the full time-dependent dynamics converges towards the cat state with good fidelity and that the cat qubit model accurately reproduces the time evolution (refer to the effective Hamiltonian in Eq. \ref{eq_effective_model}, Appendix \ref{appendix:eff_Hamiltonian}). The fidelity remaining constant over this time scale, we confirm that our DC bias approach stabilizes the cat qubit manifold without yielding more phase flips than the traditional scheme. 

\begin{table}[htbp]
\centering
\caption{Physical parameters used for numerical simulations, yielding $|g_2|/2\pi = 8.95$ MHz.}
\label{table_1}
\begin{tabular}{@{}lcl@{}}
\toprule
\textbf{Physical quantity} & \textbf{Value} \\ \midrule
$\omega_a/2\pi$            & 1.1 GHz   \\
$\omega_b/2\pi$         & 9.2 GHz  \\
$\pza$ & 0.24                      \\
$\pzb$  & 0.29  \\
$\kappa_b/2\pi$ & 
20 MHz\\
$E_J/h$        & 2.3 GHz    \\
$\omega_p/2\pi$        & 7 GHz  \\
$\omega_d/2\pi$        & 9.2 GHz\\\bottomrule
\end{tabular}
\end{table}

%% file: sections/3_noise.tex
\section{Noise}

The DC source \as{fixes a velocity for} the Josephson phase, but without providing any angular reference. It is thus essential to specifically address the impact of any noise on the \as{reference angle} for the rotating cat. Such noise can always be formulated as equivalent noise in the DC voltage source.

\begin{figure}
\includegraphics[width=\linewidth]{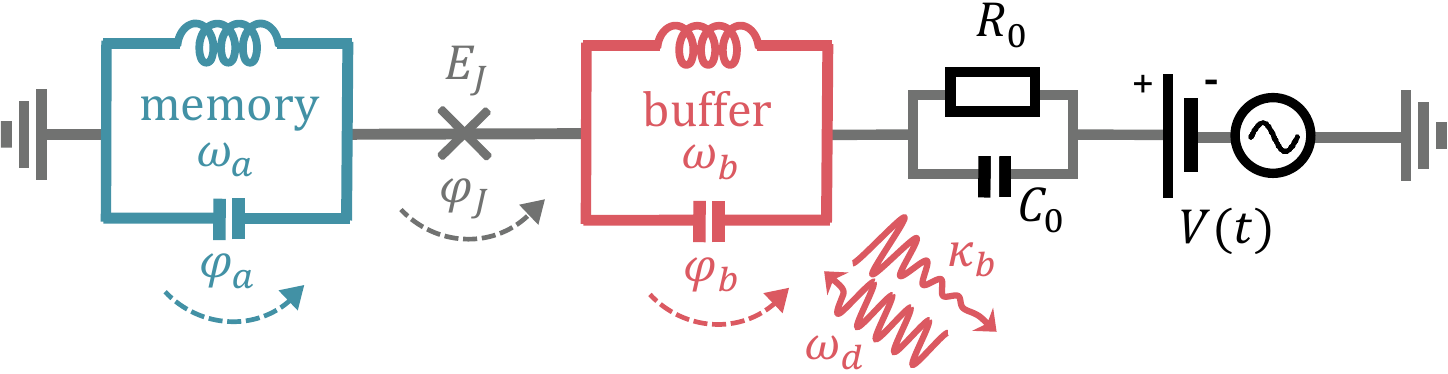}
\caption{Lumped-element model of the proposed circuit to protect against voltage noise. Injection locking requires a resistor and a microwave source at $\omega_{dc}$. Additionally, the resistor is shunted at high frequency with a capacitor.}
\label{fig:fig2a}
\end{figure}

\begin{figure}
\includegraphics[width=\linewidth]{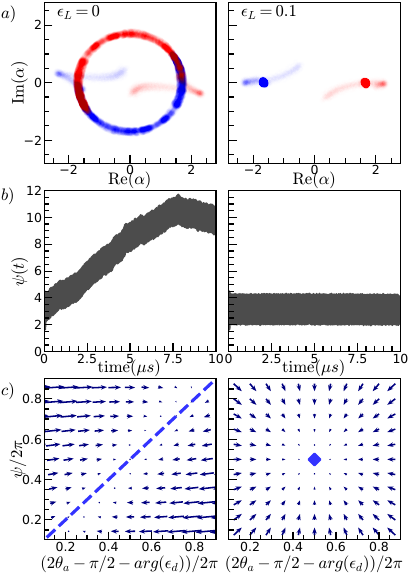}
\caption{ Classical simulations (Eq. \ref{eq:app:C5mod1} in Appendix \ref{app_injection_locking} ) illustrating cat stabilization \as{without (left: $\epsilon_L=0.0$) and with (right: $\epsilon_L=0.1$) injection locking, under a voltage noise of 206 nV rms meaning $\sqrt{\langle\delta \omega_N^2\rangle}=2\pi \times 0.1\text{GHz}$, updated independently every $dt = 0.01$ ns.} Simulation parameters are consistent with Table \ref{table_1}, with $R_0 = 100$ $\Omega$  and $C_0 = 15.9$ pF giving an RC filter cutoff frequency $\frac{1}{2\pi R_0 C0}=100$ MHz. \textbf{(a)} trajectories of $\alpha \simeq \langle  a  \rangle $ for opposite initial conditions. \as{When the locking tone provides a phase reference (right), the state converges to a pair of fixed steady states. In the absence of any locking signal (left),} the steady states' angle drifts due to voltage noise. \textbf{(b)} Evolution of $\psi(t) = \varphi_J - (\omega_b - 2\omega_a)t $, the deviation of the phase of the junction $\varphi_J$ from $\omega_{p}t$ as a function of time. The thickness of the curve is due to high frequency terms.
 With $\epsilon_L=0$ (left), $\varphi_J$ drifts freely. 
For $\epsilon_L>0$, the locking signal stabilizes $\varphi_J$ around $\pi$ (with respect to the locking signal phase reference; taking $\epsilon_L<0$  would lock $\varphi_J$ around $0$ instead, see Appendix \ref{app_injection_locking}).
\textbf{(c)} Cat state stabilization and locking stabilization represented in the ($\psi, \theta_a$) phase space. The arrows illustrate the system's evolution towards the steady state: the tail of each arrow represents the initial condition, the arrowhead points towards the steady state of the dynamics, and the length is proportional to the distance between the initial and final points. Under voltage noise (left), cat states can still be stabilized, but in the absence of a phase reference for $\varphi_J$, the angle of the cat state \as{$2\theta_a = 2\arg(\alpha)$} can take arbitrary values correlated with $\psi$ (as in \eqref{eq:argOFalpha}). By introducing the locking tone, both $\psi$ and the angle of the cat state become stabilized, resulting in a well-defined steady state solution that is independent of initial conditions (right). }
\label{fig:fig2}
\end{figure}

\paragraph*{{Voltage noise impact}}

We thus consider the experimental realisation to exhibit a finite amount of voltage noise across the junction, albeit kept low through a range of techniques \cite{Peugeot2021, Menard2022}. In particular, a voltage standard source \cite{Hamilton1997} could be used to provide extremely low intrinsic DC source noise, but practical implementation will introduce certain limitations, see Appendix \ref{appendix:standard}. 
The voltage noise $V_N(t)$ directly translates into noise in the modulating frequency, 
\begin{equation*}
\omega_{dc}(t) = \omega_p + \delta \omega_N(t)
\end{equation*}
where $\delta \omega_N(t) = 2e V_N(t) / \hbar $. This noise will impact the stabilization mechanism in several ways. 

First, the modulating frequency fluctuation can bring the system out of the frequency matching condition of Eq.~\eqref{eq:matching} which has a bandwidth $4|g_2 \alpha|^2/\kappa_b$ \cite{Lescanne2020}. 

Second (see Appendix \ref{appendix:dephasing}), 
the high-frequency components of the voltage noise imply fast random oscillations of the angle of $g_2$
in Eq.\eqref{eq:H_cat}. Equivalently, in an oscillating frame making $\epsilon_d/g_2^*$ constant, this translates into dephasing noise, with collapse operator $\sqrt{\kappa_\phi}a^\dag a$ where $\kappa_\phi$ is the effective dephasing rate that is proportional to the typical angular fluctuations. This dephasing induces no phase-flip and has a limited impact on bit-flip time provided $\kappa_\phi < 4|g_2|^2/\kappa_b$ \cite{Gautier}.

Finally, even if the first two conditions are satisfied, small fluctuations can still add up to a significant long-term drift of the angle of $g_2$ and thus of the cat angle, $\arg \alpha$. Indeed, as shown in Appendix~\ref{appendix:dephasing}, the cat angle is 
\begin{equation}\label{eq:argOFalpha}
    \theta_a = \arg \alpha(t) = \pi/4
    + \arg(\epsilon_d)/2 + \varphi_N(t)/2
\end{equation}
for slowly varying $\varphi_N(t) = \int_0^t \delta\omega_N(\tau)d\tau $. 
In other words, the angle of the cat directly depends on the integral of the voltage noise, which slowly 
drifts without bounds,
as shown in Fig~\ref{fig:fig2}. Consequently, even if the system dynamics robustly stabilize the cat qubit manifold, this implementation is unpractical since the reference angle defining the basis states will be lost after a finite time \as{--- e.g., of order $T_d$ satisfying  $\sqrt{\langle \delta\omega_N^2\rangle} dt \times \sqrt{T_d/dt} = \pi/2$ (random walk law) for $\delta\omega_N(t)$ sampled independently, with variance $\langle \delta\omega_N^2\rangle$, on intervals $dt$. Note that with this noise model, we have $\kappa_\phi \simeq
\tfrac{\pi^2}{16}\frac{1}{T_d}$ (Appendix \ref{appendix:dephasing}).}
\\

\paragraph*{{Injection locking}}
To prevent the latter effect, we propose to apply injection locking techniques to our superconducting circuit \cite{Danner2021}. 

Injection locking or forced synchronisation is extensively used to control the phase and frequency of self-sustained oscillators \cite{Pikovsky2001}, which are systems where oscillation emerges without an external periodic signal. It involves injecting a small reference signal onto which the self-oscillation will synchronize, thanks to the non-linearity and dissipation both present in the system. The injection signal can be very small, provided the frequencies of the self-sustained oscillation and injection signal are very close. When synchronisation occurs, the phase difference between the oscillator and reference remains bounded and, consequently, the frequency of the oscillation equals that of the reference on average.

A DC voltage biased junction is similar to a self-sustained oscillator: the phase progresses linearly with time under the DC voltage, yielding an oscillating current. In our system, achieving injection locking results in a constant phase shift between the junction phase and the known phase of a reference signal referred to as the locking tone whose frequency is set to $\omega_p$. Consequently, the cat angle is fixed in a known frame that rotates at frequency $\omega_a$, solving the drifting problem.

As shown in Fig.~\ref{fig:fig2a}, we thus add a microwave source to provide the locking tone and a parallel $RC$ in series to our circuit. While standard injection locking works with a bare resistance, the low-frequency $RC$ filter ensures the required dissipation for countering the slow phase drift, through the resistance $R_0$, while all the fast oscillating modes involved in the rest of the process remain unaffected thanks to the capacitive shortcut.
The total voltage bias of the system is then given by
\begin{equation}
    \frac{2eV(t)}{\hbar} = \omega_p + \delta \omega_N(t) + \epsilon_L\omega_p \cos(\omega_p t) \; ,
\end{equation}
where 
$\epsilon_L$ is the locking tone amplitude. Note that in contrast to the parametric pump case, the two-to-one photon interaction strength here does not depend on $\epsilon_L$, which can be chosen arbitrarily small as the voltage noise is reduced (see below).\\

\paragraph*{Locking performance} We analyze and simulate the system classically \cite{Massel}, an approximation that we justify for two reasons. First, both resonators have small zero point fluctuations which makes the phase close to a classical variable. Second, the quantum dynamics bring the system to superpositions of coherent states which are quasi-classical states that we reduce to their complex amplitude. For simplicity, the buffer mode is not filtered in the classical simulation, as phase flips are an effect present in the quantum simulation but are absent in the classical case.
We model the unavoidable voltage noise \as{as sampled independently at every time step $dt = 0.01$ ns,} from a Gaussian distribution with standard deviation $\sqrt{\delta \omega_N^2}/2\pi = 0.1$ GHz, which corresponds to $\delta V_{DC} = \delta \omega_N \hbar/2e  = 206$ nV. These values\as{, to leave some margin for our circuit requirements, are taken slightly higher than} those employed in state-of-the-art voltage bias experiments \cite{Peugeot2021}, \cite{Albert2024}. In the limit where $\epsilon_L \ll 1$ and $\nu_0 \ll \omega_{dc}$, with $\nu_0 = R_0/L_J = E_J R_0 (2e/\hbar)^2 $, the locking strength is characterized by the quantity $\epsilon_L \nu_0/2$ \cite{Danner2021}. Hence, once the junction energy is fixed to achieve a desired non-linearity, the locking is enabled by increasing $R_0$ or $\epsilon_L$. 
We find a satisfactory parameter regime where $R_0 = 100$ $\Omega$, $C_0 = 15.9 \,\mathrm{pF}$, \as{leading to $\nu_0 =2\pi \times 224$ MHz and RC filter cutoff frequency $\frac{1}{2\pi R_0 C_0}=$ 100 MHz}, and $\epsilon_L= 0.1$.
As shown in Fig. \ref{fig:fig2}, in the presence of the locking tone, the junction phase is locked  to the incoming signal. Some high frequency noise remains but the junction phase and hence, the cat angle, is known with respect to a reference oscillator in the setup, up to a constant that can be calibrated. \as{With the parameter values on Fig. \ref{fig:fig2}, the locking rate $\epsilon_L\nu_0/4\pi =$ 11.2 MHz is comparable to the cat qubit stabilization rate $\kappa_2/2\pi \simeq$ 16 MHz, an order of magnitude larger than the drift rate $\frac{1}{T_d} \simeq 1.6$ MHz and thus than $\kappa_\phi/2\pi \simeq 0.16$ MHz. This explains the similar, cat-stabilizing transient on the upper left and right plots. By treating the DC noise as white noise, a residual angular motion with standard deviation $\sqrt{\frac{\delta \omega_N^2 dt}{\epsilon_L \nu_0}} \simeq \pi/19$ is expected in steady state with locking (see Appendix \ref{app_injection_locking}).}

\begin{figure}
\includegraphics[width=1\linewidth]{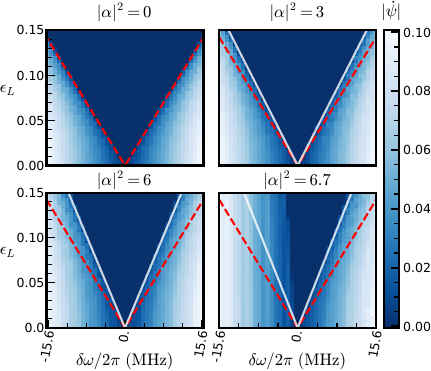}
\caption{ Arnold tongue illustrating in the $(\delta \omega, \epsilon_L)$ plane the synchronized/unsynchronized regions of the phase $\psi(t) = \varphi_J - \omega_p t$, \as{for a fixed offset $\delta\omega$}. Simulations were carried out over a duration of $t_f = 0.4 , \mu\text{s}$ using the parameters of Table \ref{table_1}, with $R_0 = 100$ $\Omega$ and $C = 15.9$ $\text{pF}$, yielding $\nu_0 = E_J R_0 (2e/\hbar)^2 = 2\pi \times 224$ MHz. The color represents the value of $\dot{\psi}$ at $t_f$. The locked region where the phase $\psi$ remains constant thanks to the external signal-induced locking (dark blue region) \as{matches the region bounded by $\epsilon_L \nu_0/2 < |\delta \omega|$ (red dashed lines) in the upper-left plot, where the memory and buffer modes are essentially empty ($|\alpha|^2=0$)}. Increasing the value of $\epsilon_d$ (from left to right) and, consequently, the population of the memory mode $|\alpha|^2$ reduces the size of the locking region, as described by the modified relation $\frac{\epsilon_L \nu_0}{2} \; |1- (\pza |\alpha|)^2| < |\delta \omega|$(white lines), discussed in Appendix \ref{appendix:photon_number_effect}. Beyond a certain photon number (here $|\alpha|^2 > 6$), this \as{first-order approximation} no longer holds, leading to an observed symmetry breaking of the locked region in the lower-right plot.} 
\label{fig:fig3}
\end{figure}

%% file: sections/4_discussion.tex
\section{Discussion}
We finally discuss the advantages and limitations of our DC cat qubit approach. 
\paragraph*{Effective junction model} 

\ta{To leverage intuition from standard superconducting circuit design, we present here an effective model for a DC-biased Josephson junction embedded in \as{a microwave circuit}. This model provides a simplified picture that is useful for understanding the behavior of the junction in \as{experiment design}.}
At first order, the averaging effect of the DC bias effectively renders the junction open: there are only fast rotating terms in Eq.~\ref{eq:dc_dev}. This is why for instance, the circuit eigenmodes are close to the bare LC-resonators: the junction acts as an open and does not couple the modes as it would in the absence of the DC voltage bias \cite{Vrajitoarea2019}.  
\as{At second order, we take into account the feedback of the oscillating signals on the junction itself.}
In fact, the DC voltage creates an oscillating current at frequency $\omega_{dc}$ with amplitude $2 e E_J/ \hbar$. If the circuit in series with the junction has a non-zero impedance $Z$ at frequency $\omega_{dc}$, then a small voltage oscillating at $\omega_{dc}$ is fed back across the junction, yielding:
\begin{equation}
    U_\mathrm{dc}(\varphi) \simeq -E_J \cos(\omega_{dc} t + \widetilde{\varphi} + \tfrac{Z}{L_J\omega_{dc}}\sin(\omega_{dc}t)) \; ,
\end{equation}
\as{with $\widetilde{\varphi}$ the remaining circuit phase. The $\tfrac{Z}{L_J\omega_{dc}}\sin(\omega_{dc}t)$ oscillation thus interferes with the DC voltage, creating non-rotating terms. In this context,} the junction then behaves as an effective small junction with energy $E_JZ/(2L_J\omega_{dc}) $ at most (depending on the complex argument of the impedance). This small effective junction slightly renormalizes the frequencies and is responsible for residual Kerr and cross-Kerr terms. For injection locking, the small voltage oscillating at $\omega_L$ \as{is applied explicitly in place of the impedance $Z$} and the effective junction creates a potential that locks the phase \cite{Danner2021}.

\paragraph*{Locking range} 

We have described two stabilization mechanisms: one inducing two stable limit cycles of the memory degree of freedom, corresponding to the cat qubit manifold, and another locking the Josephson junction phase and hence the angle of the cat qubit manifold to an external signal. These mechanisms are not independent as they originate from the same source of non-linearity, the single Josephson junction of the circuit. To investigate the mutual impact of each stabilization process, we move away from injecting noise in the system and instead consider a fixed detuning $\delta \omega$ between the effective DC bias and the locking tone. 
As shown in Fig.~\ref{fig:fig3}, for each pair of values ($\delta \omega$, $\epsilon_L$) we determine whether the system is locking or not. With $\epsilon_d=0$, we recover the well known Arnold tongue region \cite{Arnold1983} defined by $|\delta \omega|< \epsilon_L \nu_0/2$. When the memory is populated with a coherent state, the oscillation directly affects the Josephson phase, effectively narrowing the locking region. 
We find a first order correction to the width of the locking region in \as{$1-\pza^2 |\alpha|^2$ (see Appendix \ref{appendix:photon_number_effect}).}

\paragraph*{Parameter choices} 

In this paragraph, we summarize the main constraints on the parameter choices to maximize the two-to-one photon exchange rate $g_2$ between the memory and the buffer.
At first order, $g_2 \propto E_J \pza^2 \pzb$. 
Like in the previous cat qubit implementations, the buffer participation $\pzb$ is chosen large (in the limit of the RWA) as the buffer occupation remains small. The competing effects of $E_J$ and $\pza$ are better understood at a fixed value of $E_J \pza^2$, hence fixed $g_2$. When $E_J$ is increased, the first order expansion of the sine term of Eq.~\ref{eq:dc_dev}, scaling with $E_J \pza$, increases. This yields a spurious detuned drive on the cat qubit at frequency $\omega_{dc}$ which may induce bit-flips. However, higher order terms of the expansion get smaller, decreasing the occurrence of detrimental non-linear interactions. One way to counteract this linear drive is to use an additional microwave source tuned at $\omega_{dc}$ to compensate for it. Unfortunately, though this works in an ATS based stabilization, extra complications occur in the DC cat paradigm since this tone would be the locking tone.
As a side note, a large $E_J$ also favors locking by increasing $\nu_0$ (provided $\nu_0$ remains small compared to $\omega_{dc}$). 

When instead $\pza$ is increased, the system becomes more non-linear especially as the cat amplitude grows. The upper limit is defined as $2\pza \alpha < \pi$ where the junction phase oscillation due to the cat is larger than the junction potential periodicity. 
\ta{A photon number $|\alpha|^2 = 11$ should provide sufficient suppression of bit-flip errors towards fault-tolerant quantum computing \cite{Ruiz2024}. However, in order to perform detailed quantum time-dependent simulations with three modes, this work is restricted to $|\alpha|^2 = 5.5$. Beyond this value, the size of the Hilbert space made the simulations excessively demanding}. 
A possible workaround would be to consider squeezed cat states \cite{Schlegel2022, Xu2023, Hillmann2023} where equivalent bit-flip suppression can be achieved with a smaller mean amplitude. 

Regarding the system frequencies $\omega_a$, $\omega_b$ and $\omega_{dc}$, we first chose to highlight that the memory mode could be set to low frequency in order to benefit from a reduced single photon loss rate (assuming quality factors constant with frequency). This cannot be done in the autoparametric cat paradigm \cite{Marquet2023} as the buffer mode (whose frequency is constrained to twice the memory frequency) would be too low in frequency and suffer thermal excitations, detrimental for the cat qubit bit-flip time \cite{Gautier}. The buffer frequency was then chosen to avoid spurious resonances that appear when the system frequencies share low order common factors ( Appendix \ref{appendix:frequencies} ). 

%% file: sections/5_conclusion.tex
\section{Conclusion}

In this work, we have presented a new circuit design coined the DC cat qubit, which leverages a DC voltage biased Josephson junction to stabilize a cat qubit manifold. This simple circuit comprising a single Josephson junction efficiently promotes the two-to-one photon exchange interaction as the dominant term of the dynamics, significantly surpassing spurious terms such as Kerr or cross-Kerr thanks to the averaging effect of the DC voltage bias. Additionally, we have considered the effect of voltage noise on the stabilization under realistic assumptions and proposed its mitigation through injection locking.

We have simulated the system in full time-dependent dynamics, thus highlighting in an unprecedented way the effects beyond rotating wave approximation which can limit our scheme as well as others. With this, we have proposed a parameter regime that is viable for an actual experimental realization.

Beyond cat qubit stabilization, this work offers deeper insights into how DC voltage biased Josephson junctions can be used to engineer parametric interaction in superconducting circuits and highlights the critical considerations.


%% file: sections/appendix.tex

\newpage
\appendix
\onecolumngrid  

\section*{Appendix}

\as{We here give more details about the results of the main text.\newline~\newline Appendix \ref{appendix:eff_Hamiltonian} recalls the approximations involved in obtaining the (quantum) two-photon exchange Hamiltonian as a dominant contribution of our setup.\newline Appendix \ref{appendix:quantum-simulation} gives details on quantum simulation, with frequency-filtered dissipation. \newline Appendix \ref{appendix:dephasing} explains how voltage source noise can be associated to $\kappa_\phi$.\\

We next turn to the associated classical equations of motion.\newline
Appendix \ref{appendix:classical_EOM} derives the equations of motion with perfect DC source from the circuit model; then it reformulates them in terms of complex amplitudes $\alpha$ and $\beta$ akin to coherent states; and finally it performs the nonlinear classical analogue of the analysis in  Appendix \ref{appendix:eff_Hamiltonian}, confirming the bi-stable behavior of this system.\newline
Appendix \ref{app_injection_locking} recalls the higher-order classical averaging procedure (``higher-order RWA'') and applies it to a general circuit containing a Josephson Junction, in order to analyze the injection locking mechanism. It concludes with a model including voltage source noise.\newline
Appendix \ref{App:3rdorder} shows, for completeness, how the cat-stabilizing dynamics can be recovered together with the injection locking when carrying out this systematic procedure to 3rd order.\newline
Appendix \ref{appendix:photon_number_effect}, by using the insight from the previous sections, performs a smart change of frame and analyzes the interplay between locking and cat-stabilizing mechanisms.\\

We next go back to details on the quantum system.
\newline
Appendix \ref{appendix:Kerr} estimates the residual Kerr terms in our scheme, which are an order of magnitude lower thanks to using $U_{dc}$ instead of $U_p$.\newline
Appendix \ref{appendix:standard} briefly comments on using a Josephson voltage standard source.\newline
Appendix \ref{appendix:frequencies} explains our choice of oscillator frequencies to avoid spurious higher-order resonances.}

\section{Effective Hamiltonian}\label{appendix:eff_Hamiltonian}

The goal of this section is to derive the familiar effective two-photon exchange Hamiltonian\as{, assuming $\pza \alpha \,,\; \pzb \lesssim \varphi_{\text{zpf}} \ll 1$.}
We start from the time-dependent Hamiltonian in Eq. \ref{eq:H_time} of the main text, recalled here for convenience:
\begin{equation}\label{equation_ideal_sys}
\begin{split}
    H &= \hbar\omega_a a^{\dagger} a + \hbar \omega_bb^{\dagger} b  - E_J \cos \left[ \omega_{dc}t - \bm{\varphi_a} - \bm{\varphi_b}  \right] + 2\text{Re}\left(\hbar 
    \epsilon_d
    e^{-i\omega_d t}\right)(b+b^{\dagger}) \, ,
\end{split} 
\end{equation}
with $\bm{\varphi_a} = \pza (a+a^{\dagger})$ and $\bm{\varphi_b} = \pzb (b+b^{\dagger})$ and $\omega_d = \omega_b$. We can separate the even and odd terms:
\begin{equation}
\begin{split}
    H &= \hbar\omega_a a^{\dagger} a + \hbar \omega_bb^{\dagger} b + 2\text{Re}\left(\hbar{\epsilon_d}
    e^{-i\omega_d t}\right)(b+b^{\dagger})  - E_J \cos \left(\omega_{dc}t\right) \cos\left(\bm{\varphi_a} + \bm{\varphi_b}  \right)- E_J \sin \left(\omega_{dc}t\right) \sin\left(\bm{\varphi_a} + \bm{\varphi_b} \right)\, .
\end{split}
\end{equation}

The nonlinear terms can be expanded in normal order directly, after using the Baker-Campbell-Hausdorff-type formula $e^{i(v a^\dagger + v^* a)} = e^{-|v|^2/2}\, e^{i v a^\dagger} e^{i v^* a} $. We denote normal ordering of $f(.)$ by ${:}f(.){:}$  thus for instance, $e^{i(v a^\dagger + v^* a)} = e^{-|v|^2/2}\, {:}e^{i(v a^\dagger + v^* a)}{:}$ expresses the above property; as a consequence, e.g.~$E_J \, \cos\left(\bm{\varphi_a} + \bm{\varphi_b}  \right) = \tilde{E}_J \, {:}\cos\left(\bm{\varphi_a} + \bm{\varphi_b}  \right){:}$ with $\tilde{E}_J = E_J e^{-\pza^2/2-\pzb^2/2}$ as defined in the main text.
Applying a Taylor expansion up to order 3 \as{in $\varphi_{\text{zpf}}$,} we get: 
\begin{equation}
\begin{split}
    H = \hbar\omega_a a^{\dagger} a + \hbar \omega_bb^{\dagger} b + & 2\text{Re}\left(\hbar 
    \epsilon_d 
    e^{-i\omega_d t}\right)(b+b^{\dagger}) \\ & +\frac{\tilde{E}_J}{2} \cos \left(\omega_{dc}t\right) \, {:}\left(\bm{\varphi_a} + \bm{\varphi_b}  \right)^2{:} - \tilde{E}_J
    \sin \left(\omega_{dc}t\right) \Big(\vphantom{\frac{1}{3!}} \bm{\varphi_a} + \bm{\varphi_b} -\frac{1}{3!}\; {:}\left(\bm{\varphi_a} + \bm{\varphi_b}\right)^3{:} \; \Big)\, .
\end{split}
\end{equation}
\ta{The first two terms of the sine expansion are drive terms at frequency $\omega_{dc}$, i.e : $\sin ( \omega_{dc}t) \left(  \varphi_{\text{\text{zpf}},a} ( a^{\dagger} + a ) + \pzb ( b^{\dagger} + b ) \right)$. \as{We next want to perform a rotating wave approximation (RWA), whose accuracy is governed by $\Vert H \Vert / \omega$, with $\omega$ the ``rotating wave'' characteristic frequency, i.e. the smallest frequency left out as rotating. In order to achieve an RWA accuracy one order higher in $\varphi_{\text{zpf}}$, we would like to absorb these drive terms. Indeed, in the rotating frame of modes $a$ and $b$, we would then have a Hamiltonian of order $(\varphi_{\text{zpf}})^2$, meaning that the RWA would be accurate roughly up to order $(\varphi_{\text{zpf}})^4$ error terms. Note that, from the expression of $\alpha$, our assumptions imply $\epsilon_d$ of order $(\varphi_{\text{zpf}})^3$.}

The first order drive terms can be absorbed in the frame displaced by \ta{$\xi_a(t) = \xi_{a1} e^{-i\omega_{dc} t } + \xi_{a2} e^{i\omega_{dc} t }$} and  \ta{$\xi_b(t) = \xi_{b1} e^{-i\omega_{dc} t } + \xi_{b2} e^{i\omega_{dc} t }$} for $\bm{a}$ and $\bm{b}$ respectively, such that $ a \rightarrow a + \xi_a(t) $ and $ b \rightarrow b + \xi_b(t) $.
To determine the value of $\xi_a$ required to cancel the drive term $\sin(\omega_{dc}t)\pza(a+a^{\dagger})$ (focusing on mode $\bm{a}$, though the same calculation applies to mode $\bm{b}$), we can truncate the previous Hamiltonian to first order in $\pza$. This level of truncation is sufficient as it includes the terms we aim to eliminate. We write: 

\begin{equation}
    H_{trunc} = \hbar\omega_a a^{\dagger} a -\tilde{E}_J \sin(\omega_{dc} t)\left(\pzb( a + a^{\dagger}) \right)
\end{equation}
Applying the displacement described by the operator $D(-\xi_a) = e^{-\xi_a(t) a^{\dagger} + \xi_a^* a}$ which transforms $ a \rightarrow a + \xi_a(t) $ and $H_{trunc} \rightarrow D H_{trunc} D^{\dagger} + i \hbar \dot{D} {D}^{\dagger}  := \tilde{H}_{trunc}$, we get: 
\begin{equation}
    \tilde{H}_{trunc} = \hbar\omega_a (a^{\dagger} +  \xi_a^{*(t)})( a + \xi_a^{(t)}) -\tilde{E}_J \sin(\omega_{dc} t)\left(\pzb( a + a^{\dagger} + \xi_a^{(t)} + \xi_a^{*(t)})\right) + i\hbar (-\dot{\xi}_a^{(t)}a^{\dagger} + \xi_a^{*(t)}a)
\end{equation}

Now we choose $\xi_{a1}$ and $\xi_{a2}$ such that the drive term is canceled, resulting in:

\begin{equation*}
\xi_{a1} = -\frac{\varphi_{\text {zpf},a} \tilde{E}_J/(2\hbar) }{ i(\omega_{a} - \omega_{dc})}; \quad \xi_{a2} = + \frac{\varphi_{\text {zpf},a} \tilde{E}_J/(2\hbar) }{ i(\omega_{a} + \omega_{dc})}
\end{equation*}

Similarly, we perform the same calculation for mode $b$.
In this displaced frame, the Hamiltonian reads:
\begin{align}\label{eq:eqA7}
    H = & \hbar\omega_a a^{\dagger}a + \hbar\omega_b b^{\dagger}b + 2\text{Re}({\hbar \epsilon_d e^{-i\omega_d t}})(b + b^{\dagger}) \nonumber \\ + \frac{\tilde{E}_J}{2} &\cos ( \omega_{dc}t) : \left( \pza ( a^{\dagger} + a + \xi_a^{*(t)} + \xi_a^{(t)} ) + \vphantom{a^\dagger} \pzb ( b^{\dagger} + b + \xi_b^{*(t)} + \xi_b^{(t)} ) \right)^2: \nonumber \\
    + \frac{\tilde{E}_J}{3!} &\sin ( \omega_{dc}t) 
   :\left( \pza ( a^{\dagger} + a + \xi_a^{*(t)} + \xi_a^{(t)} )  
     + \vphantom{a^\dagger} \pzb ( b^{\dagger} + b + \xi_b^{*(t)} + \xi_b^{(t)} ) \right)^3:  \nonumber 
\end{align}
} 
\as{The displacement terms can be re-expressed as:
\begin{align}
    (\xi^{(t)}+ \xi^{*(t)})_{a,b} &= \bar{\xi}_{a,b} \, e^{-i\omega_{dc}t}+ \bar{\xi}_{a,b}^* \, e^{i\omega_{dc}t} \\ \nonumber
    & \text{where }\;\; \bar{\xi}_{a} = \xi_{a1}+\xi_{a2}^* \;\; \text{ and } \;\; \bar{\xi}_{b} = \xi_{b1}+\xi_{b2}^* \; .
\end{align}
}

We next go to the rotating frame at $\omega_a$ for resonator $a$ and $\omega_b$ for resonator $b$, defined by the unitary operator $U(t) = e^{i \omega_a a^{\dagger}a t} e^{i \omega_b b^{\dagger}b t}$ and resulting in :
\begin{equation}\label{eq:A7:Hrot}
\begin{aligned}
    H_{\text{rot}} &= \frac{\tilde{E}_J}{2} \cos(\omega_{dc}t)\, {:}\Big(\pza (a^\dag e^{i \omega_a t} + a e^{-i \omega_a t})    
     + \pzb (b^\dag e^{i \omega_b t} + b e^{-i \omega_b t}) + \\ & \phantom{KKKKKKKKKKKKK} 
     \as{(\pza \bar{\xi}_a+\pzb \bar{\xi}_b)\; e^{-i\omega_{dc}t} +  (\pza \bar{\xi}_a^* +\pzb \bar{\xi}_b^*)\; e^{i\omega_{dc}t}}
     \Big)^2{:}\\
    &\quad + \frac{\tilde{E}_J}{3!} \sin(\omega_{dc}t){:}\Big(\pza (a^\dag e^{i \omega_a t} + a e^{-i \omega_a t})    
     + \pzb (b^\dag e^{i \omega_b t} + b e^{-i \omega_b t}) + \\ & \phantom{KKKKKKKKKKKKK} 
     \as{(\pza \bar{\xi}_a+\pzb \bar{\xi}_b)\; e^{-i\omega_{dc}t} +  (\pza \bar{\xi}_a^* +\pzb \bar{\xi}_b^*)\; e^{i\omega_{dc}t}}
     \Big)^3{:}\\
    &\quad + 2\text{Re} (\hbar \epsilon_d 
    e^{-i\omega_d t})(b e^{-i \omega_b t} + b^\dag e^{i \omega_b t})
\end{aligned}
\end{equation}

The Rotating Wave Approximation (RWA) then consists in discarding all the terms rotating at frequencies $\omega \sim k \omega_{dc} + n\omega_a + m \omega_b$, for integers $k,n,m$, which are much faster than the Hamiltonian amplitude $\text{max}\{ (\tilde{E}_J \pza^2)/\hbar, (\tilde{E}_J \pzb^2)/\hbar, |\epsilon_d| \}$. After this, provided the frequencies are chosen to avoid spurious resonances, \as{up to order $(\varphi_{\text{zpf}})^3$ included} there only remains the static frequency combination $\omega_{dc} = +  \omega_{b} - 2\omega_{a} $ 
corresponding to the desired interaction:
\begin{equation}\label{eq:RWAinAppA}
H_{\text{eff}}/\hbar = g_2^*a^2b^{\dagger} + g_2 a^{\dagger 2} b + \epsilon_d b^{\dagger} +\epsilon_d^*b \; ,
\end{equation}
with $\hbar g_2 = \frac{\tilde{E}_J}{4i} 
\pza^2\pzb $. \as{This term is of order $(\varphi_{\text{zpf}})^3$, hence an order larger than the RWA guaranteed accuracy $\max\left( \tfrac{\varphi_{\text{zpf}}^2 E_J}{\hbar \omega_a}\; , \tfrac{\varphi_{\text{zpf}}^2 \epsilon_d}{\omega_a} \right)^2 \simeq (\varphi_{\text{zpf}})^4$, as we choose $E_J$ of the same order as $\omega_a$; and $\epsilon_d$ of order $\alpha^2 g_2$.}

We note here that the term of interest $(g_2^*a^2b^{\dagger} + h.c)$ comes from the odd sine term in $H$. The even terms are preceded by a fast oscillating $\cos(\omega_{dc}t)$, allowing to average out everything that is not close to resonant with this frequency. In contrast, the previous implementation \cite{Lescanne2020} cancels this term up to a static symmetry mismatch in the pair of Josephson junctions of their Asymmetrically Threaded SQUID element.

When adding dissipation on the buffer with $L_b = \sqrt{\kappa_b} b$, i.e.~writing
\begin{equation}
\dot\rho=-\frac{i}{\hbar}[H_{\text{eff}},\rho]+ \kappa_b
\mathcal{D} \left[b \right](\rho) \; 
\end{equation}
with $\mathcal{D} \left[b \right](\rho) = b\rho b^\dagger -\frac{1}{2} b^{\dagger} b \rho -\frac{1}{2} \rho b^{\dagger} b$,
one easily checks that the steady states of the joint system cover a two-dimensional Hilbert space spanned by $\q{\alpha_{ss}}\otimes \q{0}$ and $\q{\text{-}\alpha_{ss}}\otimes \q{0}$, with $\alpha_{ss}^2 = {-\epsilon_d}/{g_2^*}$.\\

Thanks to $\omega_{dc}$ appearing without any harmonics, avoiding spurious (quasi-)resonances is easier in the DC cat setup \as{--- see Appendix \ref{appendix:Kerr} for details}. The first unavoidable spurious terms appear in the Taylor expansion when going to 5th order in $\pza, \pzb$. \as{Performing a first-order RWA yields} the following effective Hamiltonian: 
\begin{equation}\label{eq_effective_model}
    H_{\text{eff}}/\hbar = g_2 a^{\dagger 2} b + g_2^a a^{\dagger 2} (a^\dagger a) b + g_2^b (b^{\dagger}b)\,  a^{\dagger 2} b + \epsilon_d b^{\dagger} \;\; + h.c. \; ,
\end{equation}
with $\hbar g_2^a = \frac{-\tilde{E}_J}{12i} \pza^4 \pzb =\frac{-\pza^2}{3} \hbar g_2$ and $ \hbar g_2^b = \frac{-\tilde{E}_J}{8i} \pza^2 \pzb^3 =\frac{-\pzb^2}{2} \hbar g_2$.
\as{Performing a second-order RWA pushes the RWA error to order $(\varphi_{\text{zpf}})^6$, while confirming that there are no other terms than frequency renormalizations on $\omega_a$ and $\omega_b$, up to $O(\varphi_{\text{zpf}}^5)$ included.}
The Hilbert space spanned by $\q{\alpha_{ss}}\otimes \q{0}$ and $\q{\text{-}\alpha_{ss}}\otimes \q{0}$ is preserved by the term in $g_2^b$, but not by the term in $g_2^a$. The photon parity, which corresponds to the unprotected degree of freedom of the bosonic cat-qubit code (phase flips), remains unaffected.\\

\as{The strict validity of this analysis relies on the limit $\pza\alpha\, ,\; \pzb \ll 1$, which we push rather boldly with the values taken in the main paper simulations. Nevertheless, those full-system simulations, using the fast time-varying Hamiltonian i.e.~without performing any RWA, confirm that the scheme works well.}

In fact, spurious phase-flips in our system can only originate from the limited validity of the Rotating Wave Approximation. Indeed, an uneven total power in $a$ and $a^\dagger$ in the rotating frame will always rotate with an uneven multiple of $\omega_a$, which the other frequencies in the system cannot compensate exactly \as{for a generic selection of frequencies. The Taylor series converges well even for $\pza\alpha\, ,\; \pzb$ close to or larger than $1$, thanks to its exponentially decaying prefactors for sine and cosine. The iterated RWA procedure though, i.e.~singling out exact resonances at various orders, does not converge to the true solution for any fixed $\pza\alpha\, ,\; \pzb$.}

\section{\as{Noiseless} system quantum simulation}\label{appendix:quantum-simulation}
\begin{figure}[htb]
    \centering
    \includegraphics[width=0.3\textwidth]{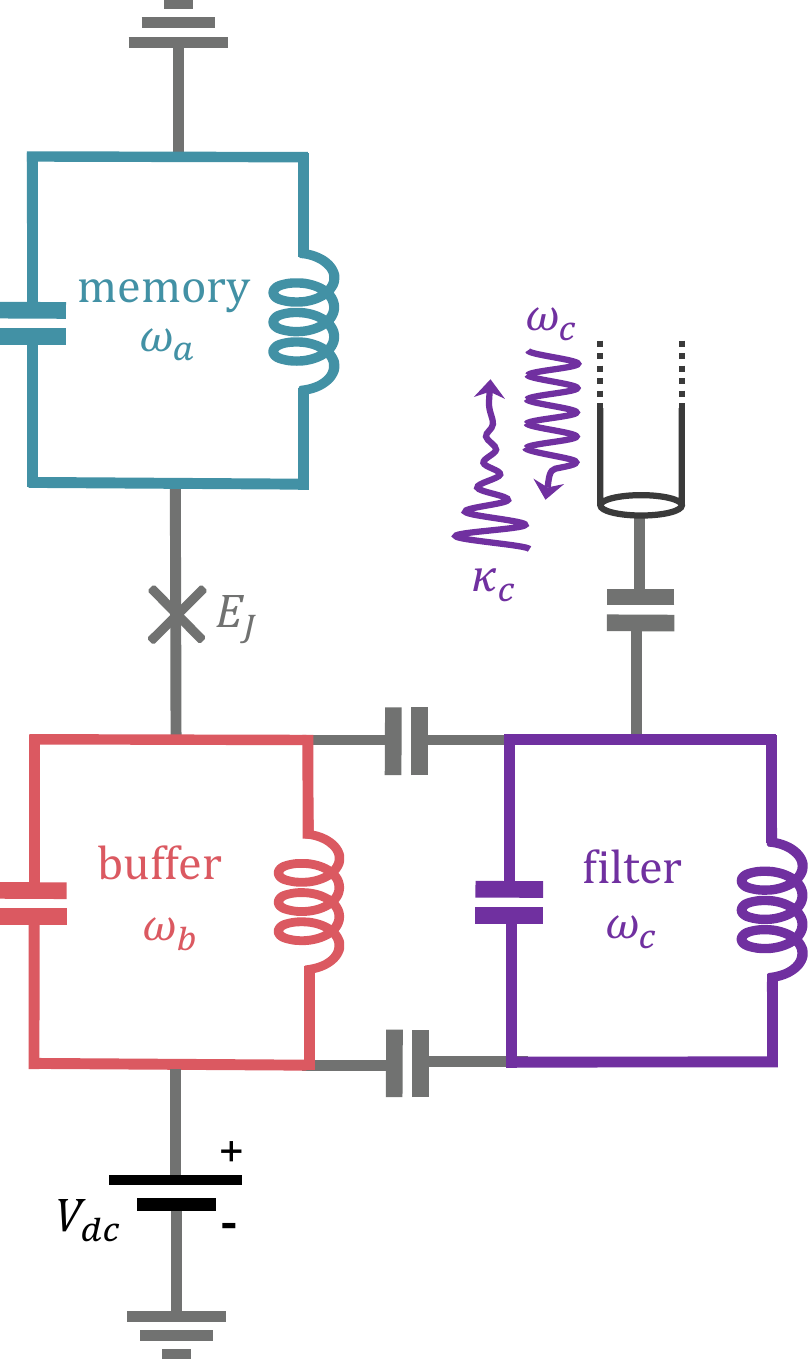} %
    \caption{Lumped-element model of the simulated DC cat qubit circuit with an additional mode $c$. This additional filter mode with a dissipation rate $\kappa_c$ is capacitively coupled to the buffer mode $b$ and protects against losses outside the frequency range centered around $\omega_c = \omega_b$.}
    \label{Fig_filter_mode}
\end{figure}
The Lindbladian equation of the noiseless system is : 
\begin{equation}\label{eq:appB:for dissfilt}
\frac{d \rho}{dt} = -\frac{i}{\hbar}\left[ H(t), \rho \right] + \mathcal{D} \left[ L_b \right](\rho) 
\end{equation}
where $L_b = \sqrt{\kappa_b} b$, and $H$ is the Hamiltonian in equation \ref{equation_ideal_sys}.\\

\paragraph*{{Filter mode}} The desired effective Hamiltonian is an approximation, considering stationary terms only, while $H$ is a time-dependent Hamiltonian. This means that modes $a$ and $b$ exchange photons, not only in pairs, but also in many ways that do not correspond to the targeted two-photon exchange mechanism. For instance, photon exchange can occur one by one through the term $\propto \cos(\omega_{dc}t) \left(a b^{\dagger} + a^{\dagger}b \right)$. \as{These undesired exchanges being oscillatory, they have a negligible net effect, as long as no other dynamics perturbs them. However, the dissipator introduced in Eq.~\eqref{eq:appB:for dissfilt} pulls mode $b$ to the vacuum, hence damping any such oscillations, and effectively inducing untolerably high phase-flip errors on mode $a$. This effect is not visible in usual cat-qubit simulations, because it is explicitly thrown out when simulating the time-averaged equations. To avoid this effect --- both in our time-resolved simulations and in experimental practice --- we must avoid damping mode $b$ at all these oscillation frequencies. For this purpose,} an additional mode, denoted $c$, is introduced (see Figure \ref{Fig_filter_mode}). This mode acts as a filter, ensuring that energy dissipation occurs primarily at the desired frequency $\omega_b$, while minimizing losses at other unwanted frequencies. The simulated system then becomes:
\begin{equation}
\frac{d \rho}{dt} = -\frac{i}{\hbar}\left[ H_f(t), \rho \right] + \mathcal{D} \left[ L_c \right](\rho) 
\end{equation}
with
\begin{equation*}\label{equation_H_filter}
\begin{split}
    H &= \hbar\omega_a a^{\dagger} a + \hbar \omega_bb^{\dagger} b + \hbar \omega_cc^{\dagger} c + g_{bc}bc^{\dagger} + g_{bc^*}b^{\dagger}c  \\ &
    - E_J \cos \left[ \omega_{dc}t - \bm{\varphi_a} - \bm{\varphi_b}  \right]  + 2\text{Re}\left(\hbar \epsilon_d
    e^{-i\omega_d t}\right)(b+b^{\dagger}) \, ,
\end{split} 
\end{equation*}
and $L_c = \sqrt{\kappa_c} c$. The resonant frequency of the filter mode $c$ is set to be identical to that of mode $b$, i.e $\omega_c = \omega_b$, while $g_{bc}$ represents the coupling strength between mode $b$ and the filter mode $c$. The relationship between the effective dissipation $\kappa_b$ and $\kappa_c$ is given by $\kappa_b \simeq \frac{4 g_{bc}^2}{\kappa_c}$ in the regime where $g_{bc} \ll \kappa_c$ \cite{Sete2015}. For the purpose of our simulation, $\kappa_c = 80$MHz and $g_{bc} = 20$MHz, which yields an effective loss rate $\kappa_b \simeq 20$MHz.\\

\paragraph*{{Simulation}} For the simulation, we have used a Qutip solver with an order 5 BDF (backward differentiation formula) integration method, for stiff differential equations. This has been necessary to cover the widely different timescales.
\as{The drive, governing the size $\alpha$ of the stabilized coherent states, was ramped up progressively in order to avoid big transients. In particular, this was meant to maintain a low occupation number in the modes $b$ and $c$, allowing for a smaller truncation of the Fock space. Figure \ref{fig:fig1} was generated with $\text{N}_a, \text{N}_b, \text{N}_c = 22, 6, 4$ where $\text{N}_i$ is the Fock basis truncation of mode $i$, and a drive amplitude linearly increased from zero to $\epsilon_d$ over the interval $t \in [0 , 5] \mu\mathrm{s}$, after which $\epsilon_d$ was held constant.}

\section{Noise-dephasing equivalence}\label{appendix:dephasing}
In this section, we want to show that in the short term, the DC voltage noise is equivalent to a dephasing dissipation channel for the cat qubit. As in the main text, we write the noisy DC source:
\begin{equation} 
    V_{DC} = \frac{\hbar}{2 e} \omega_p + V_N(t) = \frac{\hbar}{2 e} (\omega_p + \delta\omega_N(t)) .
\end{equation}
Because the phase is proportional to the derivative of the voltage $\dot{\varphi} = \frac{2e}{\hbar} V$, the initial noise on the voltage source is \textit{integrated} in time for the phase, leading to a drift over time of the Josephson Junction phase $\varphi_J$:
\begin{equation}
    \varphi_J = \omega_p\, t - \varphi_a - \varphi_b + \varphi_N(t),
\end{equation}
with $\varphi_N(t) = \int \delta\omega_N(t) \, dt$. This integration amplifies low-frequency components and attenuates high-frequency components, justifying to reconsider the RWA of Appendix \ref{appendix:eff_Hamiltonian} while treating $\varphi_N(t)$ as a quasi-static component. Instead of \eqref{eq:RWAinAppA}, we thus obtain:  
\begin{align}
    H_{\text{eff}}/\hbar &= g_2 e^{i \varphi_N(t)} a^{\dagger 2} b +  g_2^{*} e^{-i \varphi_N(t)} a^{ 2} b^{\dagger} +\epsilon_d b^{\dagger} + \epsilon_d^*b \, .
\end{align}
On a time scale where $\varphi_N(t)$ is static, when adding dissipation on the buffer, this would stabilize a cat qubit with 
$\alpha_{ss}^2 = {-\epsilon_d\, e^{i \phi_N(t)}}/{g_2^*}$, whose angle results from unknown noise. This is the third issue mentioned in the main text.

Instead, we here investigate the second issue of the main text, namely the effect of $\varphi_N(t)$ not being static, by moving to an oscillating frame with $U = e^{i \varphi_N(t) a^\dagger a /2}$. We thereby put aside for the moment the third issue, i.e.~whether this oscillating frame can be known or not to the operator. In this oscillating frame, we obtain
\begin{equation}
    \tilde{H}_{\text{eff}}/\hbar = \frac{\delta\omega_N(t)}{2} a^{\dagger} a + g_2 a^{\dagger 2}b + g_2^* a^{2}b^{\dagger} \; .
\end{equation}
Assuming that the voltage source is subject to white noise, we can write $\delta\omega_N(t) = 2 \sqrt{\kappa_\phi} w_N(t)$ such that
\begin{equation}
d\rho_t = -i g_2 [a^{\dagger 2} b + a^2 b^\dagger, \, \rho] dt + \kappa_b \mathcal{D}[b](\rho) dt \; - i \sqrt{\kappa_\phi} [a^\dagger a,\, \rho_t]\, dW_t \;
\end{equation}
follows a stochastic differential equation in It\^o sense, driven by a Wiener process 
$W_t = \int_0^t w_N(s) ds$ with expectation value $\mathbb{E}[dW_t] = 0$ and quadratic variation $\mathbb{E}[dW_t \, dW_s] = dt$ for $s=t$, zero otherwise.

Focusing on the stochastic term only, we can write the infinitesimal evolution:
\begin{eqnarray*}
    \rho_{t+dt} &=& e^{-i {\sqrt{\kappa_{\phi}}}a^{\dagger}a\, dW_t } \rho_t e^{i {\sqrt{\kappa_{\phi}}}a^{\dagger}a\, dW_t } \nonumber \\
    &=& \left( 1 -i {\sqrt{\kappa_{\phi}}} a^{\dagger}a\, dW_t -  \frac{\kappa_{\phi}}{2} (a^{\dagger}a)^2 dt  \right) \rho_t \left( 1 + i {\sqrt{\kappa_{\phi}}} a^{\dagger}a\, dW_t -  \frac{\kappa_{\phi}}{2} (a^{\dagger}a)^2 dt  \right) \nonumber \nonumber \\
    &=& \rho_t - i\sqrt{\kappa_{\phi}}a^{\dagger}a \rho_t \, dW_t + i\sqrt{\kappa_{\phi}} \rho_t a^{\dagger}a\, dW_t + \kappa_{\phi} \left( a^{\dagger}a \rho_t a^{\dagger}a - \frac{1}{2}(a^{\dagger}a)^2 \rho_t - \frac{1}{2} \rho_t (a^{\dagger}a)^2\right)dt \; .\nonumber
\end{eqnarray*}
The average state then undergoes the dynamics:
\begin{equation}\label{eq:C6dyn}
        \dot{\rho_t} = -i g_2 [a^{\dagger 2} b + a^2 b^\dagger, \, \rho] + \kappa_b \mathcal{D}[b](\rho) + \kappa_{\phi} \mathcal{D}[a^{\dagger}a](\rho) \; ,
\end{equation}
which thus adds standard dephasing dissipation to the target dynamics. The rate $\kappa_\phi$ corresponds to white noise of amplitude $\frac{\hbar}{2e}\sqrt{\kappa_\phi}$ in $V_N(t)$.

\as{Strictly speaking, Eq.~\eqref{eq:C6dyn} expresses the evolution undergone by the state in the oscillating frame, if we had no knowledge at all of the Wiener process $W_t$ (and hence of the relation $U(t)$ between the oscillating frame and lab frame). In practice, when the low-frequency drift of $U(t)$ is countered by injection locking, the $\kappa_\phi$ expressed here can be understood as the residual effect of all the medium- and high-frequency noise.}\\

\as{Regarding noise modeling, a Wiener process (integral of white noise) is the mathematical description of Brownian motion. Its essential feature here is that, starting with a known position at $t=0$, the position at time $t$ is a Gaussian with standard deviation $\sqrt{t}$. Considering $\delta\omega_N$ to follow a white noise signal --- being totally uncorrelated among infinitesimally differing times --- is a standard but abstract idealization. Our simulations all consider a somewhat less unphysical situation, where $\delta\omega_N$ is sampled independently on finite intervals $dt=0.01$ns. After a large number of steps $N=t/dt$, the corresponding Brownian motion then diffuses as a Gaussian with standard deviation $\sqrt{N}\sigma$, with $\sigma$ the standard deviation for diffusion over the interval $dt$. The latter corresponds to the standard deviation of $\delta\omega_N$, integrated over $dt$. Thus altogether, we get a standard deviation
$$\sigma_t = \sqrt{\langle \delta\omega_N^2\rangle} dt \times \sqrt{t/dt}$$ 
at time $t$, as expressed in the main text. 

The white noise and simulated noise can be related to each other by equalling the resulting standard deviations once $t \gg dt$, we get:
\begin{equation}\label{eq:whitenoiseequiv}
\text{equivalent white noise amplitude }= \sqrt{\langle \delta\omega_N^2\rangle} \sqrt{dt} \; .
\end{equation}
With this, we get $\kappa_\phi = \frac{dt \langle \delta\omega_N^2\rangle}{4} = \tfrac{\pi^2}{16}\frac{1}{T_d}$ with $T_d$ as defined in the main text.}

\section{Classical equations of motion with perfect DC source}\label{appendix:classical_EOM}

The scheme depicted on Figure \ref{fig:fig1}.b is governed by the following equations:
\begin{equation}\label{eq:app:C1}
\begin{cases}
    \begin{aligned}
        &\ddot{\varphi}_a  = -\omega_a^2 \varphi_a +  \frac{2e}{\hbar}\frac{I_J}{C_a}\sin\varphi_J\\
        &\ddot{\varphi}_b = -\omega_b^2 \varphi_b - {\kappa_b} \dot{\varphi}_b + \frac{2e}{\hbar}\frac{I_J}{C_b}\sin\varphi_J +\text{Re}({r_d e^{-i\omega_d t}})\\
        & \dot{\varphi}_J =  \frac{2e}{\hbar} V_{\text{dc}} - \dot{\varphi}_a -\dot{\varphi}_b \; .
    \end{aligned}
\end{cases}
\end{equation}
The first equation expresses the conservation of current (times $\frac{2e}{\hbar C_a}$) at the node connecting the $a$ oscillator to the Josephson junction (JJ), with $\dot{\varphi}_a=\frac{2e}{\hbar}V_a$, $V_a$ the voltage across the oscillator and $\omega_a = 1/\sqrt{L_aC_a}$. The first two terms express the current through the capacitor and inductance respectively. The last term expresses the AC Josephson effect: the actual current in the JJ is proportional to $I_J \sin(\varphi_J)$, and would thus oscillate if $\phi_J$ increases linearly for a constant voltage $V_J$. The JJ critical current $I_J = \frac{2e}{\hbar}E_J$ is linked to the Josephson energy $E_J$. The second equation is similar for the $b$ oscillator, adding to it a drive of amplitude $r_d$ at frequency $\omega_d$, and dissipation similar to Ohm's law (i.e.~proportional to $\kappa_b \dot{\varphi}_b \sim V_b/R$). Those additional elements are not represented as such on the figure because they are rather implemented via a transmission line setup. Finally, the third equation covers voltage conservation around the circuit.
After defining $E_C = \frac{4e^2}{8C}$, $\omega_{\text{dc}}=\frac{2e}{\hbar} V_{\text{dc}}$, $E_L=\frac{1}{L}\Big(\frac{\hbar}{2e} \Big)^2$ and the variables 
\begin{equation}
n_a = \frac{\hbar}{8 E_{Ca}}\dot{\varphi}_a \quad , \quad 
n_b = \frac{\hbar}{8 E_{Cb}}\dot{\varphi}_b \; ,
\end{equation}
the previous equations can be put in standard form: 
\begin{equation}
\begin{cases}
    \begin{aligned}
        &\dot{\varphi}_a = \frac{8E_{Ca}}{\hbar} n_a \\
        &\dot{n}_a  = -\frac{E_{La}}{\hbar} \varphi_a + \frac{E_J}{\hbar}\sin\varphi_J \\
        &\dot{\varphi}_b = \frac{8E_{Cb}}{\hbar} n_b \\
        &\dot{n}_b  = -\frac{E_{Lb}}{\hbar} \varphi_b  + \frac{E_J}{\hbar}\sin\varphi_J - {\kappa_b} n_b + \frac{\hbar}{8E_{Cb}} \text{Re}({r_d e^{-i\omega_d t}})  \\
        & \dot{\varphi}_J = \omega_{\text{dc}} - \frac{8 E_{Ca}}{\hbar} n_a - \frac{8 E_{Cb}}{\hbar} n_b  \; .
    \end{aligned}
\end{cases}
\end{equation}
We can now introduce the zero point fluctuations as follows : 
\begin{align}
    \tilde{\varphi} = \frac{\varphi}{\varphi_{\text{zpf}}};\; \tilde{n} = \frac{n}{n_{\text{zpf}}} \quad \text{ with } \quad \varphi_{\text{zpf}} = \left( \frac{2E_C}{E_L} \right)^{1/4};\; n_{\text{zpf}} = \frac{1}{2}\left( \frac{E_L}{2E_C} \right)^{1/4} \; .
\end{align}
Using these two equations, we finally get : 
\begin{equation}\label{eq:app:C5}
\begin{cases}
    \begin{aligned}
        \dot{\tilde{\varphi}}_a &= \omega_a \tilde{n}_a \\
        \dot{\tilde{n}}_a  &= -\omega_a \tilde{\varphi}_a + 2\frac{E_J}{\hbar}\pza\sin\varphi_J\\
        \dot{\tilde{\varphi}}_b &= \omega_b \tilde{n}_b \\
        \dot{\tilde{n}}_b  &= -\omega_b \tilde{\varphi}_b + 2\frac{E_J}{\hbar}\pzb\sin\varphi_J - {\kappa_b} \tilde{n}_b + \frac{1}{\pzb\omega_b}\text{Re}({r_d e^{-i\omega_d t}}) \\
        \dot{\varphi}_J &= \omega_{\text{dc}} - \pza \omega_a\, \tilde{n}_a -\pzb \omega_b\, \tilde{n}_b \, .
    \end{aligned}
\end{cases}
\end{equation}
By defining 
\begin{equation}
\alpha = \frac{1}{2}\left({\tilde{\varphi}_a} + i {\tilde{n}_a}\right) \quad , \quad \beta = \frac{1}{2}\left({\tilde{\varphi}_b} + i {\tilde{n}_b}\right) \; ,
\end{equation}
we get:
\begin{equation}\label{eq:classmod:alphabeta}
\begin{cases}
    \begin{aligned}
        \dot{\alpha} &= -i\omega_a \alpha  \ta{+}i\frac{E_J}{\hbar}\pza \sin\left(\varphi_J \right)\\ 
        \dot{\beta} &= -i\omega_b \beta  \ta{+}i\frac{E_J}{\hbar} \pzb \sin\left(\varphi_J \right) - i{\kappa_b} \text{Im}(\beta)- 2i \text{Re}{(\epsilon_d e^{-i\omega_d t})}\\
        \dot{\varphi}_J &= \omega_{dc} -  2\pza \omega_a \text{Im}(\alpha) - 2\pzb \omega_b \text{Im}(\beta) \; ,
\end{aligned}
\end{cases}
\end{equation}
with $\epsilon_d = -r_d/(4 \pzb\omega_b)$. 

Note that the dissipation $\kappa_b$ only affects one quadrature. Its usual expression, inducing effective damping of the amplitude $\beta$, will be retrieved after performing RWA in a frame rotating at the oscillator frequency. Also note that, independently of the other circuit elements, we have  $\frac{d\text{Re}(\alpha)}{dt} = \omega_a\, \text{Im}(\alpha)$ and $\frac{d\text{Re}(\beta)}{dt} = \omega_b\, \text{Im}(\beta)$, which will be useful in further computations.\\[3mm]

Before adding the injection locking mechanism, let us analyze the dynamics \eqref{eq:classmod:alphabeta} as such. The third equation can be readily integrated to get:
\begin{equation}\label{eq:appAS:phiJ}
    \varphi_J = \varphi_J(0)+\omega_{dc} t - 2 \pza \text{Re}(\alpha) - 2 \pzb \text{Re}(\beta) \; .
\end{equation}
We can already observe that any drift in $\varphi_J(0)+\omega_{dc} t$ will just be kept, leading to an arbitrary reference angle in absence of further dynamics. Resolving this issue will be the subject of injection locking, see next Section. For now, we assume $\varphi_J(0)=0$ and concentrate on the remaining equations.

Plugging in \eqref{eq:appAS:phiJ}, the dynamics of the two cavities becomes:
\begin{equation}\label{eq:C9}
\begin{cases}
    \begin{aligned}
        \dot{\alpha} =& -i\omega_a \alpha  \ta{+}i\frac{E_J}{\hbar}\pza \sin\left(\omega_{dc} t - \pza (\alpha+\alpha^*) - \pzb (\beta+\beta^*) \right)\\ 
        \dot{\beta} =& -i\omega_b \beta  \ta {+}i\frac{E_J}{\hbar} \pzb \sin\left(\omega_{dc} t - \pza (\alpha+\alpha^*) - \pzb (\beta+\beta^*) \right) \\ &- \frac{\kappa_b}{2}(\beta-\beta^*)- 2i \text{Re}{(\epsilon_d e^{-i\omega_d t})} \; .
\end{aligned}
\end{cases}
\end{equation}
We next pursue along the same lines as in Appendix \ref{appendix:eff_Hamiltonian}. We separate the sine into terms with $\sin(\omega_{dc} t)$ and $\cos(\omega_{dc} t)$. By going to \as{the displaced frame with $\xi_a , \xi_b$ and} rotating frame at $\omega_a$ for $\alpha$ and $\omega_b$ for $\beta$, i.e.~defining $\tilde{\alpha} = \as{(\alpha -\xi_a^{(t)})}\, e^{i\omega_a t}$ and $\tilde{\beta} = \as{(\beta-\xi_b^{(t)})}\, e^{i\omega_b t}$, we get:
\begin{equation}\label{eq:newD10}
    \begin{cases}
    \begin{aligned}
        \dot{\tilde{\alpha}} =& \; i\frac{E_J}{\hbar}\pza e^{i\omega_a t}  \sin(\omega_{dc} t) \cos_1\Big( - \pza (\tilde\alpha e^{-i\omega_a t}+\tilde\alpha^*e^{i\omega_a t}) - \pzb (\tilde\beta e^{-i\omega_b t}+\tilde\beta^* e^{i\omega_b t}) \\
        & \phantom{KKKKKKKKKKK} - \as{(\pza \bar{\xi}_a+\pzb \bar{\xi}_b)\; e^{-i\omega_{dc}t} +  (\pza \bar{\xi}_a^* +\pzb \bar{\xi}_b^*)\; e^{i\omega_{dc}t}}\Big)\\ 
        & + \; i\frac{E_J}{\hbar}\pza e^{i\omega_a t}  \cos(\omega_{dc} t) \sin\Big( - \pza (\tilde\alpha e^{-i\omega_a t}+\tilde\alpha^*e^{i\omega_a t}) - \pzb (\tilde\beta e^{-i\omega_b t}+\tilde\beta^* e^{i\omega_b t}) \\
        & \phantom{KKKKKKKKKKK} - \as{(\pza \bar{\xi}_a+\pzb \bar{\xi}_b)\; e^{-i\omega_{dc}t} +  (\pza \bar{\xi}_a^* +\pzb \bar{\xi}_b^*)\; e^{i\omega_{dc}t}}\Big)\\ 
        \dot{\tilde{\beta}} =&  \; i\frac{E_J}{\hbar} \pzb e^{i\omega_b t} \sin(\omega_{dc} t) \cos_1\Big( - \pza (\tilde\alpha e^{-i\omega_a t}+\tilde\alpha^*e^{i\omega_a t}) - \pzb (\tilde\beta e^{-i\omega_b t}+\tilde\beta^* e^{i\omega_b t}) \\
        & \phantom{KKKKKKKKKKK} - \as{(\pza \bar{\xi}_a+\pzb \bar{\xi}_b)\; e^{-i\omega_{dc}t} +  (\pza \bar{\xi}_a^* +\pzb \bar{\xi}_b^*)\; e^{i\omega_{dc}t}}\Big)\\
        & + i\frac{E_J}{\hbar} \pzb e^{i\omega_b t} \cos(\omega_{dc} t) \sin\Big( - \pza (\tilde\alpha e^{-i\omega_a t}+\tilde\alpha^*e^{i\omega_a t}) - \pzb (\tilde\beta e^{-i\omega_b t}+\tilde\beta^* e^{i\omega_b t}) \\
        & \phantom{KKKKKKKKKKK} - \as{(\pza \bar{\xi}_a+\pzb \bar{\xi}_b)\; e^{-i\omega_{dc}t} +  (\pza \bar{\xi}_a^* +\pzb \bar{\xi}_b^*)\; e^{i\omega_{dc}t}}\Big)\\ &- \frac{\kappa_b}{2}(\tilde\beta-\tilde\beta^* e^{2i\omega_b t}))- i (\epsilon_d e^{i(\omega_b-\omega_d) t} + \epsilon_d^* e^{i(\omega_b+\omega_d) t}) \; ,
\end{aligned}
\end{cases}
\end{equation}
\as{where $\cos_1(x) = \cos(x)-1$.}
The time-dependent equation on the right-hand side can then be approximated by its time-average (Rotating Wave Approximation, RWA), provided its oscillating terms move much faster than $\tilde{\alpha}$ and $\tilde{\beta}$. We must thus select $\frac{E_J}{\hbar}\varphi_{\text{zpf}}, \kappa_b, \epsilon_d$ much smaller than $\omega_a,\omega_b,\omega_{dc},\omega_d$, and in principle than all their integer combinations appearing when expanding the sine,cosine of $\alpha e^{i\omega_a t}$ and $\beta e^{i\omega_b t}$. Alternatively, we Taylor expand the sine,cosine to finite order and consider higher powers in $\varphi_{\text{zpf}}$ as a perturbation, not further specified, to be countered by the stabilizing dynamics found at lower order. \as{More precisely, thanks to the change of frame, and by taking into account how $\kappa_b,\epsilon_d$ are selected a posteriori (namely both comparable to $g_2$ which will be of order 3), the right-hand side of Eq.\eqref{eq:newD10} is of order $2$ in $\frac{E_J}{\hbar}\varphi_{\text{zpf}}$. Then the result of the RWA, i.e.~just keeping non-rotating terms, will be accurate up to an error of order 4 in $\frac{E_J}{\hbar}\varphi_{\text{zpf}}$.}

We thus consider the resonance conditions $\omega_d = \omega_b$ and $\omega_{dc} = \omega_b - 2 \omega_a$ and assume that we avoid any other resonances at low order. \as{The expansion of $\sin()$ contains terms rotating with an uneven multiple of either $\omega_a$ or $\omega_b$. Considering the non-rotating terms, up to and including order $(\varphi_{\text{zpf}})^3$, there only remains:}
\begin{equation}\label{eq:eqA5}
\begin{cases}
\begin{aligned}
    \dot{\alpha} &\simeq  -2 i  g_2\, {\alpha}^{*}  {\beta} \\
    \dot{\beta} &\simeq -ig_2^* \,{\alpha}^2 - \frac{\kappa_b}{ 2} {\beta} - i \epsilon_d \; ,
\end{aligned}
\end{cases}
\end{equation}
with $\hbar g_2 = \frac{E_J}{4i} 
\pza^2\pzb $ defined as in the quantum case.
This system features three steady states.
\begin{itemize}
\item Equilibrium at $\alpha_{ss}=0$ and $\beta_{ss}=r$ with $r =-\frac{2 i \epsilon_d}{\kappa_b}$: By linearizing the dynamics around this equilibrium, we obtain: 
$$\dot{\alpha} = -2 i g_2 r \, \alpha^* \quad , \quad \dot{\beta} = - \tfrac{\kappa_b}{2} (\beta-r) \; .$$
This equilibrium is unstable, since an equation of the type $\dot{z}(t) = c\, z^*(t)$ is unstable for any nonzero $c \in \mathbb{C}$.
\item A pair of equilibria at 
\begin{equation}\label{eq:app:clacccat}
\beta_{ss}=0 \quad , \quad (\alpha_{ss})^2=\frac{-\epsilon_d}{g_2^*} \; .
\end{equation}
Linearizing around these, we get
$$\dot{\alpha} = -2 i g_2 \alpha_{ss}^* \, \beta \quad , \quad \dot{\beta} = -2 i g_2^* \alpha_{ss} (\alpha-\alpha_{ss}) - \tfrac{\kappa_b}{2} \, \beta \; .$$
This linear system has a sum of eigenvalues $-\kappa_b/2 < 0$ and a product of eigenvalues $4|g_2|^2 |\alpha_{ss}|^2 > 0$ and is thus stable indeed, at least locally. More precisely, the local eigenvalues are $\frac{-\kappa_b}{4} \pm \frac{\kappa_b}{4}\sqrt{1-\frac{4 |g_2|^2 |\alpha_{ss}|^2}{(\kappa_b/4)^2}}$, which yield $(\frac{-\kappa_b}{2}\,,\; - 2 \kappa_2 |\alpha_{ss}|^2)$ when $\kappa_b \gg \kappa_2 |\alpha_{ss}|^2$. This is the classical equivalent of the well-known dissipative cat-qubit stabilization.
\end{itemize}

\as{Note that, in absence of any stabilization on the integral $\varphi_J$, the phase of $g_2$ here takes an arbitrary value, depending on the arbitrary initial condition $\varphi_J(0)$. We here take $\varphi_J(0)=0$, but we will later see that our injection locking signal rather stabilizes $\varphi_J(0)=\pi$, leading to a sign change on $g_2$.}

\section{Classical dynamics with filtered injection locking}\label{app_injection_locking}

We now introduce the injection locking mechanism, see e.g. references \cite{Balanov}, \cite{Pikovsky2001}.

Standard injection locking works by adding a periodic signal, close to the free rotation frequency, in series with a resistance dissipating the deviations from a locking situation. Since any dissipating element bears the danger of corrupting quantum information, we want to restrict the one ensuring injection locking, to the minimal necessary frequency domain. The primary role of injection locking is to avoid \emph{long-term drift} of the cat qubit angle, so it should be active at low frequencies, thus ideally in a quantum context, at low frequencies only. For this reason, we would replace the standard locking resistance by a dynamic element, e.g.~a parallel RC circuit. We y have to show that the injection locking mechanism works in combination with the cat stabilization dynamics.

We therefore start with a general target circuit and a general locking filter, on which we will impose constraints as we go. The DC cat qubit scheme with an RC filter serves as a guiding example; but we thus want to highlight that there is flexibility in the actual implementation. The general linear circuit can be described in state space form
\begin{equation}\label{eq:app:StateSpace1}
    \dot{z} = \hat{A} z  + \hat{B}_J \sin(\phi_J) + \hat{B}_d(t) \; ,
\end{equation}
where matrix $\hat{A}$ describes the circuit intrinsic dynamics, vector $\hat{B}_d(t)$ the input drives and vector $\hat{B}_J$ how it is affected by the input \emph{current} from the Josephson Junction (JJ). This circuit in turn contribues an additional voltage to the loop containing the JJ, expressed as
\begin{equation}\label{eq:app:StateSpace2}
    \dot{\varphi}_J = \omega_{dc} + \hat{C}_J \, z + D_J \sin(\phi_J) + D_d(t) \; .
\end{equation}
Here $D_d(t)$ contains external AC voltage signals (e.g.~the ``locking tone''), while the linear circuit's voltage contribution is put in standard state space output form $y=C z + D u$ (with $z$ the state, $y$ the output, $u$ the input). One checks that this form works at least for every linear circuit which contains a path without inductances; otherwise, the JJ nonlinearity appears at more places. The direct feedthrough term $D_J$ vanishes if the circuit contains a path with capacitors only.\\

\subsection{Analyzing the dynamics with higher-order RWA (averaging theory)}\label{ref:AppERWA}

Injection locking is a long-term stabilization phenomenon in a fast oscillating system, thus featuring a timescale separation $\varepsilon \ll 1$. As such, locking can usually be highlighted by performing a time-dependent change of frame, approximately following the dominant oscillations, in order to push residual oscillations to higher orders in $\varepsilon$ and focus instead on the secular evolution. Although several approaches are possible, we here follow the systematic procedure of ``averaging'' from the mathematical systems theory literature \cite{sanders2007averaging}. With this, we obtain the locking effect in 4 steps as a ``second-order RWA'' phenomenon. For each step, we recall the general averaging theory and exemplify it on several use cases with injection locking: a general linear circuit; a standard locking circuit alone, consisting of just a locking signal and a resistance in series with the Josephson Junction (JJ); a locking circuit replacing the resistance by a parallel RC; and our cat-qubit case with locking signal and parallel RC. 

\as{In order to treat all these various cases, we perform no system-specific change of frame to the equations here. In Appendix \ref{appendix:eff_Hamiltonian}, the system-specific displacement allowed us to push the initial dynamics to order $(\varphi_{\text{zpf}})^2$ insteaf of order $\varphi_{\text{zpf}}$, such that a single order of averaging allowed us to draw conslusions up to $O(\varphi_{\text{zpf}})^4$ errors. We will get back to such approach in Appendix \ref{appendix:photon_number_effect} and Appendix \ref{appendix:Kerr}, after getting a better insight on the system behavior in presence of locking.}

Note that the general averaging approach can be used to study any long-term effects other than locking.\\

\underline{1. The starting point is a dynamical system in standard form:} 
$$\dot{x}=f(x,t) \, $$ where $f(\cdot, t)$ oscillates at frequencies $\omega$ larger than some characteristic frequency $\omega_c$, and $\Vert f \Vert \, /\, \omega_c = \varepsilon \ll 1$ i.e.~the system motion itself is much slower than the characteristic oscillation frequency. We assume all functions to be smooth, in particular with derivatives in $x$ being $O(1)$.\\

$\bullet$ For the general linear circuit \eqref{eq:app:StateSpace2}, the principle is that the DC bias dominates $\dot{\varphi}_J$, assuming $\omega_c$ of the same order as $\omega_{dc}$. We must thus remove the latter from the dynamics. It may be not be known exactly, since this is not the frequency of an external signal, but results from the voltage bias. Hence, we will use a signal of known frequency $\omega_L$ and define $\varphi_J \rightarrow \hat{\varphi}_J = \varphi_J - \omega_L t$  with $\delta\omega = \omega_{dc}-\omega_{L} \ll \omega_c$. Following the same principle, we further assume $C_J, D_d, D_J \ll \omega_c$.

The linear target circuit can be put in the requested form, with motion $\Vert f \Vert$ much slower than $\omega_c$, when described in an appropriate time-dependent frame. We denote the dynamics before this change of frame with $\hat{}\;$. We only have to assume $\Vert B_J \Vert \ll \omega_c$, i.e.~the nonlinear effect of $\varphi_J$ is not big enough to significantly modify the dominant frequencies. When $\hat{B}_d(t) \neq 0$, we can write the dynamics in a time-dependent \emph{displaced} frame that approximately follows the corresponding motion of $z$, such that the residual $\Vert B_d(t) \Vert \ll \omega_c$. Furthermore, the circuit intrinsic dynamics $\hat{A}\, z$ would typically contain eigenvalues with frequencies (i.e.~imaginary parts) of the same order as $\omega_{dc}$. These can be covered by going to a \emph{rotating} frame, expressed by a unitary $U(t)=\exp(-i \Omega t)$ with the Hermitian $\Omega$ expressing the fast rotation rate. Then $\dot{z}$ in this rotating frame replaces $\hat{A}$ by $A(t) = U(t) (\hat{A}-i\Omega) U(t)^\dagger =: U(t) \hat{A}_0 U(t)^\dagger$. We assume that any large contributions to $\hat{A}$ feature imaginary eigenvalues, such that $\hat{A}_0$ would have eigenvalues of order $\ll \omega_c$. Concretely, this essentially means that we assume all damping rates in $\hat{A}$ to be $\ll \omega_c$.  This assumption is natural if we want to speak of well-resolved frequencies. 

These changes of frame on the target circuit modify the stationary and oscillating parts of $f(x,t)$. In particular, the rotating frame introduces oscillations in $A(t) = U(t) \hat{A}_0 U(t)^\dagger$,  $B_J(t) = U(t) \hat{B}_J$, $C_J(t) = \hat{C}_J U(t)^\dagger$ which would be absent from the bare modeling in lab frame. It also modifies the typical frequencies in $B_d(t) = U(t) \hat{B}_d(t)$. The frame displacement reduces the amplitude of $B_d(t)$, but adds terms of this frequency to $D_d(t)$ (and to $\delta\omega$ if $\hat{B}_d(t)$ had a constant term); a small point of attention is to maintain the latter $\ll \omega_c$. In practice, the displacement may often be unnecessary, namely when the drive $\Vert \hat{B}_d(t) \Vert$ would naturally be comparable to the circuit damping rates, hence $\ll \omega_c$. We thus assume a model:
\begin{eqnarray}\label{eq:E2}
\dot{\hat{\varphi}}_J &=& \delta\omega + C_J(t)\, z + D_d(t) + D_J \sin(\hat\varphi_J + \omega_L t) \\ \nonumber
\dot{z} &=& A(t)\, z + B_J(t)\, \sin(\hat\varphi_J + \omega_L t) + B_d(t) \; ,
\end{eqnarray}
in which all terms are $\ll \omega_c$.\\ 

$\bullet$ For the standard resistance-based locking circuit alone, the voltage on the resistance is directly proportional to the current, so we have
\begin{eqnarray}\label{eq:E-Ronly3} 
            \dot{\hat{\varphi}}_J &=& \delta \omega - \nu_0  \sin(\hat{\varphi}_J + \omega_L t ) + \epsilon_L\omega_L\cos(\omega_L t)\; ,
\end{eqnarray}
with $\nu_0 = \frac{2e}{\hbar} I_J R$ and $\epsilon_L$ the locking tone amplitude. Thus $z$ has dimension $0$, we only have $D_d(t)$ the locking tone and $D_J=-\nu_0$. The averaging conditions are saisfied if $\epsilon_L \ll 1$ and $\nu_0 \ll \omega_L$.\\

$\bullet$ For the RC locking circuit alone, we thus have a parallel RC in series with the JJ and a voltage composed of the DC bias and a tone at $\omega_L$. The variable $z$ comports a single component reflecting the RC dynamics, for instance $z=\tau \, \frac{2e}{\hbar} \, V_R$ with $V_R$ the voltage across the RC element and $\tau=RC$ its characteristic time, compared to which low frequencies are damped by the resistance while high frequencies are shunted through the capacitor. The dynamics then writes:
\begin{eqnarray}\label{eq:E3} 
            \dot{\hat{\varphi}}_J &=& \delta \omega - z/\tau + \epsilon_L\omega_L\cos(\omega_L t)\\ \nonumber
            \dot{z}  &=&  - z/\tau + \nu_0  \sin(\hat{\varphi}_J + \omega_L t ) \; ,
\end{eqnarray}
with $\nu_0 = \frac{2e}{\hbar} I_J R$ and $\epsilon_L$ the locking tone amplitude. The locking tone and resistance, meant to counter small drifts, are usually chosen very small, i.e.~$\epsilon_L \ll 1$ and $\nu_0 \ll \omega_L$. Furthermore,  we impose $\frac{1}{\tau} \ll \omega_L$, i.e.~focusing on an ambitious RC filter, for which only the frequencies well below $\omega_L$ undergo a resistive effect. Under these conditions, \eqref{eq:E3} is already in an appropriate form for averaging, without the need for a displacement or rotating frame. More precisely, it takes the form \eqref{eq:E2} with
$C_J(t)=-1/\tau$, $D_d(t)=\epsilon_L\omega_L\cos(\omega_L t)$, $A(t)=-1/\tau$, $B_J(t) = \nu_0$ and $D_J=B_d(t)=0$.\\

$\bullet$ For the cat-qubit example, we go back to Section \ref{appendix:classical_EOM} and add the RC locking mechanism in series. Calling $n_R=\tau \, \frac{2e}{\hbar} \, V_R$ the associated variable, this transforms \eqref{eq:app:C5} into
\begin{equation}\label{eq:app:C5mod1}
\begin{cases}
    \begin{aligned}
        \dot{\hat{\varphi}}_a &= \omega_a \hat{n}_a \\
        \dot{\hat{n}}_a  &= -\omega_a \hat{\varphi}_a + \tfrac{2E_J\pza}{\hbar}\sin(\hat{\varphi}_J+\omega_L t)\\
        \dot{\hat{\varphi}}_b &= \omega_b \hat{n}_b \\
        \dot{\hat{n}}_b  &= -\omega_b \hat{\varphi}_b + \frac{2E_J\pzb}{\hbar}\sin(\hat{\varphi}_J+\omega_L t) - {\kappa_b} \hat{n}_b - 4\text{Re}({\epsilon_d e^{-i\omega_d t}}) \\
        \dot{n}_R &= -n_R/\tau + \nu_0 \sin(\hat{\varphi}_J+\omega_L t)\\
        \dot{\hat{\varphi}}_J &= \delta\omega - \pza \omega_a\, \hat{n}_a -\pzb \omega_b\, \hat{n}_b - n_R / \tau + \epsilon_L \omega_L \cos(\omega_L t) \, .
    \end{aligned}
\end{cases}
\end{equation}
Like in Section \ref{appendix:classical_EOM}, we must go to a rotating frame for the two oscillators $a$ and $b$. In the above coordinates, the frame change writes $U(t) = \text{blockdiag}[R_{\omega_a t} \; ;\;  R_{\omega_b t} \; ;\; 1]$ with $R_{\theta} = [\cos\theta,\; -\sin\theta \, ; \, \sin\theta,\; \cos\theta\,]$. After the coordinate change, plugging in $\omega_d=\omega_b$ with $\epsilon_d$ real, we obtain:
\begin{equation}\label{eq:app:C5mod2}
\begin{cases}
    \begin{aligned}
        \dot{\varphi}_a &= \frac{-2E_J\pza}{\hbar}\sin(\hat{\varphi}_J+\omega_L t)\, \sin(\omega_a t) \\
        \dot{n}_a  &= \frac{2E_J\pza}{\hbar}\sin(\hat{\varphi}_J+\omega_L t)\, \cos(\omega_a t)\\
        \dot{\varphi}_b &= \frac{-2E_J\pzb}{\hbar}\sin(\hat{\varphi}_J+\omega_L t)\, \sin(\omega_b t) + 4\epsilon_d \sin(\omega_b t)\cos(\omega_b t)  - \sin^2(\omega_b t) \, \kappa_b\, \varphi_b + \sin(\omega_b t)\cos(\omega_b t) \, \kappa_b \, n_b\\
        \dot{n}_b  &=  \frac{2E_J\pzb}{\hbar}\sin(\hat{\varphi}_J+\omega_L t)\, \cos(\omega_b t) - 4\epsilon_d \, \cos^2(\omega_b t)  - \cos^2(\omega_b t) \, \kappa_b\, n_b + \sin(\omega_b t)\cos(\omega_b t) \, \kappa_b \, \varphi_b \\
        \dot{n}_R &= -n_R/\tau + \nu_0 \sin(\hat{\varphi}_J+\omega_L t)\\
        \dot{\hat{\varphi}}_J &= \delta\omega - \pza \omega_a\, (\cos(\omega_a t)n_a\text{-}\sin(\omega_a t)\varphi_a) -\pzb \omega_b\, (\cos(\omega_b t)n_b\text{-}\sin(\omega_b t)\varphi_b) - n_R / \tau + \epsilon_L \omega_L \cos(\omega_L t) \, .
    \end{aligned}
\end{cases}
\end{equation}
This equation is in the form \eqref{eq:E2} provided $\varphi_{\text{zpf}}, |\epsilon_L| \ll 1$ and $\delta\omega,1/\tau,\nu_0,\kappa_b,\varphi_{\text{zpf}}E_J/\hbar,|\epsilon_d| \ll \omega_L$. These have all been assumed and motivated above. In the form \eqref{eq:E2}, we have $D_J=0$ and $D_d(t) = \epsilon_L \omega_L \cos(\omega_L t)$; row vector $C_J$ and column vector $B_J$ have one frequency per component, respectively $\omega_a,\omega_a,\omega_b,\omega_b,0$ in the order of the coordinates; $B_d(t)$ contains nonzero components only for $(n_b,\varphi_b)$, featuring frequencies $0$ and $2\omega_b$; and finally, $A(t)$ is block-diagonal with a zero block on $(n_a,\varphi_a)$, a block in $\kappa_b$ with frequencies $0$ and $2\omega_b$ on $(n_b,\varphi_b)$; and $-1/\tau$ on $n_R$.\\[3mm]

\underline{2. The first-order average dynamics is obtained} by just taking the time-average of $f(x,t)$ assuming $x$ fixed. In other words, we define 
\begin{equation}\label{eq:appav:dotxsep}
\dot{x} = \bar{f}(x) + \tfrac{\partial}{\partial t} \tilde{f}(x,t)
\end{equation}
where $\bar{f}$ has no explicit time-dependence and $\tfrac{\partial}{\partial t} \tilde{f}(\cdot,t)$ has zero time-average; at first order, 
$$\dot{x} \simeq \bar{f}(x) \, .$$
Usually, the locking phenomenon does not appear yet at this first order. Writing $\tfrac{\partial}{\partial t} \tilde{f}(\cdot,t)$ with a derivative just facilitates notation in later steps. We will also choose $\tilde{f}(\cdot,t)$ with zero time-average. Note that, with $\Vert f \Vert / \omega_c = \varepsilon \ll 1$ and all oscillations $\omega \geq \omega_c$, we have $\Vert \tilde{f} \Vert =O(\varepsilon)$.\\

$\bullet$ For the general linear circuit, the time-averaged dynamics boils down to
\begin{eqnarray*}
\dot{\hat{\varphi}}_J & \simeq & \bar{C}_J\, z + \delta\omega \\
\dot{z} &\simeq& \bar{A}\, z + {\overline{B_J(t)\, \sin(\hat\varphi_J + \omega_L t)}} +  \bar{B}_d \; ,
\end{eqnarray*}
with the overline denoting time-average. There are two cases.
\begin{itemize}
    \item[-] Often, the linear system and hence $B_J(t)= U(t) \hat{B}_J$ features no component directly at frequency $\omega_L$. Namely, it is avoided to have this direct linear resonance, because the $\omega_L$ bias is precisely meant to trigger a nonlinear parametric resonance. In this case, the term $\overline{B_J(t)\, \sin(\hat\varphi_J + \omega_L t)}$ drops and the right hand side of the differential equations is independent of $\hat\varphi_J$. This reflects marginal stability, i.e.~$\hat\varphi_J$ can freely drift under noises in e.g.~$\delta\omega$. In other words, besides $\bar{A}$ which is assumed to have no eigenvalues with positive real part, the system features an eigenvalue $0$ for $\hat\varphi_J$. It thus makes sense to examine the behavior at higher order.
    \item[-] If $B_J(t)$ does feature a frequency $\omega_L$, then $\overline{B_J(t)\, \sin(\hat\varphi_J + \omega_L t)}$ is a linear combination of components of the type $\frac{B_{\ell}}{2}\cos(\hat\varphi_J-b_{\ell})$. Due to the nonlinearity, depending on the parameter values, the system may or may not have some steady states, and depending on the Jacobian the latter may or may not be stable. This stability would dominate whatever is found at the next order, unless eigenvalues have zero real part.
\end{itemize}
Going on, we assume to be in the first case, i.e.~$\overline{B_J(t)\, \sin(\hat\varphi_J + \omega_L t)}=0$. Regarding the oscillating part, we define $\tilde{S}_J(t)$ and $\tilde{Q}_J(t)$ such that $B_J(t) \sin(\hat\varphi_J+\omega_L t) = \sin(\hat\varphi_J) \, \tfrac{\partial}{\partial t} \tilde{S}_J - \cos(\hat\varphi_J)\, \tfrac{\partial}{\partial t} \tilde{Q}_J$. We also define e.g.~$\tilde{A}$ such that $A(t) = \bar{A}+\tfrac{\partial}{\partial t} \tilde{A}(t)$ and similarly for $C_J(t),\, D_d(t),\, B_{d}(t)$. Note that $\tilde{D}_d(t)$ is then just the primitive of $D_d(t)$. We may further observe that $\bar{C}_J \, z = \hat{C}_J \,\overline{U(t)^\dagger} \, z$ will only contain the non-oscillating parts of the linear target circuit.\\

$\bullet$ For the R locking circuit alone, the first order reduces to $\dot{\varphi}_J = 0$ and there is nothing more to say.\\

$\bullet$ For the RC locking circuit alone, we have $B_J(t)=\nu_0$ constant and hence we are in the first case indeed. The first-order average dynamics becomes:
\begin{eqnarray*}
    \dot{\hat{\varphi}}_J &\simeq& -z/\tau + \delta\omega \\
    \dot{z} &\simeq& -z/\tau \; ,
\end{eqnarray*}
while $\tilde{S}_J = \frac{\nu_0}{\omega_L}\sin(\omega_L t)$ and $\tilde{Q}_J = \frac{\nu_0}{\omega_L}\cos(\omega_L t)$.\\

$\bullet$ In the cat-qubit example, we have chosen to avoid spurious resonances, hence $B_J(t)$ features $\omega_a,\omega_b \neq \omega_L$ and we are again in the first case. The average dynamics at first order becomes:
\begin{eqnarray*}
    \dot{\hat{\varphi}}_J \;\simeq\; -n_R/\tau + \delta\omega &\qquad,\qquad& \dot{n}_R  \;\simeq\; -n_R/\tau \\
    \dot{n}_a  \;\simeq\; 0 &\qquad,\qquad& \dot{n}_b \;\simeq\; \tfrac{-\kappa_b}{2}\, n_b - 2 \epsilon_d \\
    \dot{\varphi}_a  \;\simeq\; 0 &\qquad,\qquad& \dot{\varphi}_b \;\simeq\;  \tfrac{-\kappa_b}{2}\, \varphi_b \; .    
\end{eqnarray*}
At this order, the dynamics on the $RC$ element converges to $0$, while the whole dynamics is independent of $\varphi_J$. The oscillator $a$ does not move at this order, while the oscillator $b$ converges to the first steady state of \eqref{eq:eqA5} (which turned out to be unstable there, once the oscillator $a$ dynamics become conclusive). We also have:
$$
\tilde{S}_J = \left[\begin{array}{l} 
\tfrac{E_J\pza}{\hbar}\left(\tfrac{\cos((\omega_a+\omega_L)t)}{\omega_a+\omega_L}+\tfrac{\cos((\omega_a-\omega_L)t)}{\omega_a-\omega_L}\right)\\
\tfrac{E_J\pza}{\hbar}\left(\tfrac{\sin((\omega_a+\omega_L)t)}{\omega_a+\omega_L}+\tfrac{\sin((\omega_a-\omega_L)t)}{\omega_a-\omega_L}\right)\\
\tfrac{E_J\pzb}{\hbar}\left(\tfrac{\cos((\omega_b+\omega_L)t)}{\omega_b+\omega_L}+\tfrac{\cos((\omega_b-\omega_L)t)}{\omega_b-\omega_L}\right)\\
\tfrac{E_J\pzb}{\hbar}\left(\tfrac{\sin((\omega_b+\omega_L)t)}{\omega_b+\omega_L}+\tfrac{\sin((\omega_b-\omega_L)t)}{\omega_b-\omega_L}\right)\\
\tfrac{\nu_0}{\omega_L} \sin(\omega_L t)
\end{array}\right] \qquad , \qquad \tilde{Q}_J = \left[\begin{array}{l} 
\tfrac{-E_J\pza}{\hbar}\left(\tfrac{\sin((\omega_a+\omega_L)t)}{\omega_a+\omega_L}-\tfrac{\sin((\omega_a-\omega_L)t)}{\omega_a-\omega_L}\right)\\
\tfrac{E_J\pza}{\hbar}\left(\tfrac{\cos((\omega_a+\omega_L)t)}{\omega_a+\omega_L}-\tfrac{\cos((\omega_a-\omega_L)t)}{\omega_a-\omega_L}\right)\\
\tfrac{-E_J\pzb}{\hbar}\left(\tfrac{\sin((\omega_b+\omega_L)t)}{\omega_b+\omega_L}-\tfrac{\sin((\omega_b-\omega_L)t)}{\omega_b-\omega_L}\right)\\
\tfrac{E_J\pzb}{\hbar}\left(\tfrac{\cos((\omega_b+\omega_L)t)}{\omega_b+\omega_L}-\tfrac{\cos((\omega_b-\omega_L)t)}{\omega_b-\omega_L}\right)\\
\tfrac{\nu_0}{\omega_L} \cos(\omega_L t)
\end{array}\right] \; .
$$
\\[3mm]

\underline{3. To push the oscillating vector field to the next order}, we perform the change of variable
\begin{equation}\label{eq:app:AvCF1}
x_1 = x - \tilde{f}(x,t) \; ,
\end{equation}
which is close to identity and hence well-defined, since $\Vert \tilde{f} \Vert = O(\varepsilon)$. Replacing this in \eqref{eq:appav:dotxsep}, we get:
\begin{eqnarray*}
    \dot{x}_1 &=& \bar{f}(x) - \nabla_x \tilde{f}(x,t) \cdot \Big( \bar{f}(x) + \tfrac{\partial}{\partial t} \tilde{f}(x,t) \Big)\\
    &=& \bar{f}(x_1) \;\; + \nabla_{x_1}\bar{f} \cdot \tilde{f}(x_1,t) \;\; - \nabla_{x_1} \tilde{f} \cdot \Big(\bar{f}(x_1) + \tfrac{\partial}{\partial t} \tilde{f}(x_1,t)\Big) \quad + O(\varepsilon^3) \; ,
\end{eqnarray*}
where in the second line we have inverted \eqref{eq:app:AvCF1} approximately as $x = x_1 + \tilde{f}(x_1,t) + O(\varepsilon^2)$ and performed a limited Taylor expansion.
By keeping only the nonoscillating terms up to second order included, we obtain
$$\dot{x}_1 \simeq \bar{f}(x_1)\;\; - \overline{\nabla_{x_1}\tilde{f} \cdot \tfrac{\partial}{\partial t} \tilde{f}(x_1,t)} $$
where the overline denotes time-average.\\

$\bullet$ For the general linear circuit, the change of variable writes:
\begin{eqnarray*}
    \varphi_{J,1} &=& \hat{\varphi}_J - \tilde{C}_J \, z - \tilde{D}_d + \tfrac{D_J}{\omega_L} \cos(\hat\varphi_J + \omega_L t) \\
    z_1 &=& z - \tilde{A}\, z  + \cos(\hat\varphi_J)\, \tilde{Q}_J - \sin(\hat\varphi_J) \, \tilde{S}_J - \tilde{B}_d \; .
\end{eqnarray*}
The corresponding non-oscillating dynamics, up to second order included, can be simplified by recalling the $U(t)$ rotating frame. Writing $U(t) = \bar{U}+\tfrac{\partial}{\partial t} \tilde{U}(t)$, we have $\tilde{U}=i U(t)\Omega^{-1}$ with $\Omega^{-1}$ the pseudo-inverse of $\Omega$. Then we have for instance $\bar{C}_J = \hat{C}_J \bar{U}^\dagger$ and $\tilde{C}_J = -i \hat{C}_J \Omega^{-1} U(t)^\dagger$. Further, we assume that  $D_d(t)$ contains a component $D_L \sin(\omega_L t + d_L)$ and that $U(t)$ contains no component at frequency $\omega_L$ (see point 2.). Under these conditions, the resulting average dynamics can be written as:
\begin{eqnarray*}
    \dot{\varphi}_{J,1} &\simeq & \hat{C}_J \, (I+i\Omega^{-1} \hat{A}_0)\, \bar{U}^\dagger\, z_1 \;\;\; - \frac{D_J D_L}{2\omega_L} \cos(\varphi_{J,1}-d_L) \\
    && + \delta\omega - \frac{D_J^2}{2\omega_L} + i\hat{C}_J \Omega^{-1}\overline{\hat{B}_d} \; ,\\[3mm]
    \dot{z}_1 & \simeq & \left(\bar{A}-\overline{\tilde{A}\tfrac{\partial\tilde{A}}{\partial t}} \right)\, z_1 \;\;\; - \overline{\tilde{S}_J D_d}\, \cos(\varphi_{J,1}) - \overline{\tilde{Q}_J D_d}\, \sin(\varphi_{J,1}) \\ 
    && - D_J \left( \overline{\tilde{S}_J\sin(\omega_L t)}\, \cos^2(\varphi_{J,1}) + \overline{\tilde{Q}_J\cos(\omega_L t)}\, \sin^2(\varphi_{J,1}) +
    (\overline{\tilde{S}_J\cos(\omega_L t)+\tilde{Q}_J \sin(\omega_L t)})\, \cos(\varphi_{J,1})\sin(\varphi_{J,1})\right)\\
    && - \overline{\tilde{S}_J C_J}\, z_1\, \cos(\varphi_{J,1}) - \overline{\tilde{Q}_J C_J}\, z_1\, \sin(\varphi_{J,1}) \;\;\; + \bar{B}_d - \overline{\tilde{A} B_d}  \; .
\end{eqnarray*}
 We recall that $\hat{A}_0$ here denotes the dynamics before going to rotating frame, but \emph{already subtracting the fast dynamics that is rotated out.}
Of course, some of those terms can drop depending on the frequency arrangement. For instance, the second line of $\dot{z}_1$ reduces to a constant if $U(t)$ contains no frequency $2\omega_L$, because its only non-oscillating terms would then come from $\tilde{S}_J = \sin(\omega_L t) M_0$ and $\tilde{Q}_J = \cos(\omega_L t) M_0 $ for some $M_0$.\\

$\bullet$ For the R locking circuit alone, given the locking tone $D_d(t) = D_L \sin(\omega_L t + d_L)$ with $D_L = \epsilon_L \omega_L$ and $d_L=\pi/2$, we get:
$$\dot{\varphi}_{J,1} = \delta\omega - \tfrac{\nu_0^2}{2\omega_L} + \tfrac{\epsilon_L\nu_0}{2}\, \sin(\varphi_{J,1})\; .$$
\\

$\bullet$ For the RC locking circuit alone, we have $\bar{U}=I$ and $\Omega=\tilde{A}=0$ (no rotating frame). Using the computed expressions for $\tilde{S}_J$ and $\tilde{Q}_J$, the dynamics reduce to:
\begin{eqnarray*}
    \dot{\varphi}_{J,1} &\simeq& -z_1/\tau + \delta\omega \\
    \dot{z}_1 &\simeq& -z_1/\tau -\tfrac{\epsilon_L \nu_0}{2}\, \sin(\varphi_{J,1}) \; .
\end{eqnarray*}\\

$\bullet$ In the cat-qubit example $\Omega^{-1}=\text{blockdiag}[\;\tfrac{-1}{\omega_a} \sigma_y\, ;\, \tfrac{-1}{\omega_b} \sigma_y \;;\; 0\;]$ with $\sigma_y = [0,\; -i \, ; \, i,\; 0\,]$. With this, $\tilde{A}$ involves a nonzero block only on the oscillator $b$, which induces some second-order corrections to its dynamics. The further matrices and vectors in lab frame $\hat{}$ can be identified from \eqref{eq:app:C5mod1}, the ones in the rotated frame from \eqref{eq:app:C5mod2}. We recall that $D_J=0$. In the terms with $\tilde{S}_J,\tilde{Q}_J$, all the contributions on oscillators $a$ and $b$ vanish because they involve 3-wave mixing, while the resonance condition \eqref{eq:matching} is at 4-wave mixing. There thus remains little more than the oscillator $b$ dynamics from the first order, plus the locking dynamics:
\begin{eqnarray*}
    \dot{\varphi}_{J,1} \;\simeq\; -n_R/\tau + \delta\omega &\qquad,\qquad& \dot{n}_R  \;\simeq\; -n_R/\tau - \tfrac{\epsilon_L \nu_0}{2}\, \sin(\varphi_{J,1}) \\
    \dot{n}_a  \;\simeq\; 0 &\qquad,\qquad& \dot{n}_b \;\simeq\; \tfrac{-\kappa_b}{2}\, n_b + \tfrac{\kappa_b}{2} \tfrac{\kappa_b}{4\omega_b}\varphi_b - 2 \epsilon_d \\
    \dot{\varphi}_a  \;\simeq\; 0 &\qquad,\qquad& \dot{\varphi}_b \;\simeq\;  \tfrac{-\kappa_b}{2}\, \varphi_b - \tfrac{\kappa_b}{2} \tfrac{\kappa_b}{4\omega_b}n_b - 2 \epsilon_d \tfrac{\kappa_b}{4\omega_b}\; .    
\end{eqnarray*}
To obtain nontrivial dynamics on oscillator $a$ and see the classical equivalent of cat state stabilization \eqref{eq:app:clacccat} with this approach, we would have to go one order further in the averaging, i.e.~performing a change of variable which pushes the oscillating vector field to the third order. See Appendix \ref{App:3rdorder}.\\[3mm]

\underline{4. The last point is to analyze} the second-order dynamics, where locking is expected to appear. If an exponentially stable behavior is found at this order, then (for sufficiently small $\varepsilon$) it will dominate and persist in the full dynamics.\\

$\bullet$ The general linear circuit can be partly discussed on the basis of its frequency content.
\begin{itemize}
    \item[-] A stationary component is needed in the circuit to possibly stabilize the JJ, i.e.~to have $\dot{\varphi}_{J,1}$ depend on the variables. Indeed, $\dot{\varphi}_{J,1}$ can directly depend on $z_1$, only if the circuit has a non-rotating component in $\bar{U}$ (corrected to 2nd order, if this non-rotating component, through $\hat{A}_0$, is coupled to a rotating one); or, it can directly depend on $\varphi_{J,1}$ when $D_J\neq 0$, which we recall appears when the circuit features a component not shunted by a capacitor. In other words, the state of any rotating modes has no direct influence on $\dot{\varphi}_{J,1}$. All this is consistent with the aim of injection locking, to avoid long-term drift.
    \item[-] The circuit $\dot{z}_1$ first depends on its intrinsic dynamics with $\bar{A}$ at 1st order, corrected by possible internal resonances at 2nd order. Assuming that the chosen $D_d(t)$ contains the locking tone only, $\dot{z}_1$ further depends on $\phi_{J,1}$ through elements of $U(t)$ which are either stationary or at frequency $2\omega_L$. In the last line, a joint dependence on $z_1$ and $\varphi_{J,1}$ appears only if a sum or difference of two frequencies in $U(t)$ is resonant with $\omega_L$.
    \item[-] Thus altogether, a first somewhat general scenario could consider $\bar{U}=0$ i.e.~only rotating modes. Then $\dot{\varphi}_{J,1}$ does not depend on $z_1$. We thus need $D_J\neq 0$ i.e.~a series resistance to stabilize the JJ with a direct dependence on $\varphi_{J,1}$. We then have a pair of steady state solutions, e.g.~$\bar{\varphi}_{J,1}$ and $\pi-\bar{\varphi}_{J,1}$ when choosing $d_L=\pi/2$, provided $\tfrac{D_J D_L}{2\omega_L}$ is larger than the sum of biases; in particular, $\delta\omega$ must be of second order. Typically, one solution is unstable and one is stable. This $\bar{\varphi}_{J,1}$ then appears as a given in the equations of $\dot{z}_1$ governing the linear circuit.
    \item[-] Another somewhat general scenario could consider oscillatory modes in series with a non-rotating ``filter'' dynamics, through which $\bar{U}\neq 0$. In this case, $\dot{\varphi}_{J,1}$ depends on $z_1$. If the filter features a capacitive shunt, we further have $D_J=0$, so the stability of $\dot{\varphi}_{J,1}$ is determined solely by its coupling to filter part of $z_1$. If the biased JJ was seeking to activate resonances of the type $\omega_k = 2\omega_L$ or $\omega_j \pm \omega_k = \omega_L$, with $\omega_j,\omega_k$ some linear circuit modes, then these would interfere with the locking process. Otherwise, we have a system of the form
    $$\dot{\varphi}_{J,1} = C_s z_1 + b_s \quad , \quad  \dot{z}_1 = A_r\, z_1 +b_r + B_s \, \sin\varphi_{J,1} \; ,$$
    where $C_s$ and $B_s$ only involve the non-rotating ``filter'' modes of $z_1$, and assuming the phase $d_L=\pi/2$ in the locking signal is chosen to ensure $\overline{\tilde{S}_J D_d}=0$. Now, if the linear circuit itself features a valid steady state with $A_r,b_r$, and $A_r$ is invertible on the non-rotating modes, then with $D_L$ chosen not too small the full system always has a steady state too. Out of the two steady-state values $\bar{\varphi}_{J,1}$ and $\pi-\bar{\varphi}_{J,1}$, typically one will be stable and one unstable. However, this cannot be guaranteed as firmly as in the first case without studying the full circuit.\\
\end{itemize}

$\bullet$ For the R locking circuit, note the additional bias in $\tfrac{\nu_0^2}{2\omega_L}$: at second order, the natural junction dynamics is thus shifted. Provided $|\tfrac{\epsilon_L\nu_0}{2}| > |\delta\omega - \tfrac{\nu_0^2}{2\omega_L}|$, there are two steady states $\bar{\varphi}_{J,1}$. By linearizing the dynamics in the vicinity of the steady states, one obtains the convergence rate $\tfrac{\varepsilon_L\nu_0\cos(\bar{\varphi}_{J,1})}{2}$ which must be negative for the stable equilibrium. For $\delta\omega - \tfrac{\nu_0^2}{2\omega_L} \sim 0$, the junction can lock around $\bar{\varphi}_{J,1} \sim 0,\pi$. Taking $\epsilon_L<0$ would stabilize $\bar{\varphi}_{J,1}=0$, i.e.~matching the arbitrary phase choice made in \eqref{eq:argOFalpha} and Appendix~\ref{appendix:dephasing} before discussing the phase reference. Taking $\epsilon_L>0$ instead stabilizes a phase $\bar{\varphi}_{J,1}=\pi$ with respect to this arbitrary choice.\\

$\bullet$ For the RC locking circuit alone, a steady state corresponds to $-z_1/\tau = \tfrac{\epsilon_L\nu_0}{2} \sin(\varphi_{J,1})$. Replacing this in the equation for $\dot{\varphi}_{J,1}$ yields the same steady state as for the R locking circuit alone, except there is no bias on $\delta\omega$. The stability condition is also the same. In fact, for $1/\tau \gg \epsilon_L\nu_0$ , the eigenvalues of the dynamics linearized around the steady state are $-1/\tau$ and $\tfrac{\varepsilon_L\nu_0\cos(\bar{\varphi}_{J,1})}{2}$. The first one is the RC filter characteristic time, proper to the linear circuit itself; the second one is the same convergence rate as in the standard R locking circuit. For $\delta\omega \sim 0$, the junction can lock around $\bar{\varphi}_{J,1} \sim 0,\pi$. Again, taking $\epsilon_L<0$ would stabilize $\bar{\varphi}_{J,1}=0$, matching the arbitrary phase choice made in \eqref{eq:argOFalpha} and in Appendix~\ref{appendix:dephasing}, before discussing the phase reference. Taking $\epsilon_L>0$ instead stabilizes a phase $\bar{\varphi}_{J,1}=\pi$ with respect to this arbitrary choice. The latter choice explains the shift by $\pi$ of both the horizontal and vertical axis on the bottom right panels of Figure \ref{fig:fig2}(b).\\

$\bullet$ In the cat-qubit example, at this order, the dynamics on $\varphi_{J,1}$ and $n_R$ is totally uncoupled from the two oscillators. Furthermore, the cat stabilization process, which involves a matching condition between 3 frequencies, is not yet visible at this 2nd order of averaging, where only products of 2 oscillating terms are computed.\\

\underline{In further steps, the procedure can be repeated} to get the average evolution to higher order accuracy, namely including products of several oscillating terms. In fact, by proceeding to the next order, the cat-stabilization dynamics can be retrieved in this classical model, see Appendix \ref{App:3rdorder}.

For this, the first requirement is to track the change of variables more accurately in Step 3. By inverting
$x_1 = x- \tilde{f}(x)$ with 
\begin{equation}\label{eq:app:TaylExp}
x = x_1+\tilde{f}(x_1+\tilde{f}(x_1+...))
\end{equation}
and carrying out all Taylor expansions to sufficient order, the associated approximation error can be pushed to high order. Next, we write the resulting dynamics for $x_1$ as in Step 2, namely: 
\begin{equation}\label{app:eq:for3rd}
\dot{x}_1 = \bar{f}_1(x_1)+\tfrac{\partial}{\partial t} \tilde{f}_1(x_1,t) \; , \end{equation}
but where now $\tfrac{\partial}{\partial t} \tilde{f}_1$ is of order two in the timescale separation. A change of variable analogous to Step 3 brings the time-dependent vector field to order three. The procedure can then be repeated to even higher orders, until one decides to discard the remaining rotating terms. It is important to note that, with quasi-periodic systems, in many cases this iteration will have to be stopped at an order where quasi-resonances appear \footnote{Indeed, these would make $\tilde{f}_k$ much larger than $\tfrac{\partial}{\partial t} \tilde{f}_k$, to a point that $\epsilon^k$ with $\epsilon \ll 1$ could not compensate.}.

\as{For a system that stabilizes in the vicinity of a steady state, another approach consists in performing a change of frame after which the residual dynamics is of order $\varepsilon^k$ with $k>1$. For instance, with $k=2$, first-order averaging (Step 2.) gives a residual error $O(\varepsilon^4)$ and second-order averaging (Step.3) gives a residual error $O(\varepsilon^6)$. This allows a faster refinement of the system analysis, as carried out in Appendix \ref{appendix:photon_number_effect}.}

\as{\subsection{Including noise in the model}\label{app:ssec:OUnoise}

Echoing the analysis of Appendix \ref{appendix:dephasing}, we can write a stochastic model including white noise and analyze its behavior in presence of the injection locking dynamics. From the analysis just made, we can reduce this to the simple model :
\begin{equation}\label{eq:stoch:Rlocking}
    d\varphi_{J,1}^{(t)} = \tfrac{\epsilon_L\nu_0}{2} \sin(\varphi_{J,1}^{(t)}) \, dt + 2 \sqrt{\kappa_\phi}\, dW_t \;\; \simeq \;\; \tfrac{-\epsilon_L\nu_0}{2} \, \varphi_{J,1}^{(t)} \, dt + 2 \sqrt{\kappa_\phi}\, dW_t \; ,
\end{equation}
assuming $\delta\omega_N(t)$ to be white noise associated to a Wiener process $dW_t$ and with amplitude $2\sqrt{\kappa_\phi}$, see Appendix \ref{appendix:dephasing}, and $\varphi_{J,1}$ close to $\pi$. The last from of Eq.\eqref{eq:stoch:Rlocking} is known as an Ornstein-Uhlenbeck process. Under this linear stochastic equation, the probability distribution of $\varphi_{J,1}^{(t)}$ converges towards a Gaussian steady state, with standard deviation $\frac{2\sqrt{\kappa_\phi}}{\sqrt{\epsilon_L \nu_0}}$. For the noise process simulated in the paper, considering the equivalence \eqref{eq:whitenoiseequiv} yields the steady-state standard deviation $\sqrt{\frac{\delta\omega_N^2 \, dt}{\epsilon_L \nu_0}}$ reported in the main text.}

\section{Averaging the classical dynamics to 3rd order} 
\label{App:3rdorder}

This section shows how, by carrying out the systematic averaging procedure to 3rd order on the classical dynamics, we can recover the cat-stabilizing equations in presence of injection locking. This approach also avoids the necessity of carrying out a preliminary change of variables: instead of obtaining $O(\epsilon^3)$ accuracy by performing 1st-order RWA on dynamics of magnitude $O(\epsilon^2)$, we obtain it with 3rd-order RWA on dynamics of magnitude $O(\epsilon)$. The main text draws no particular conclusions from this, so we include it more for completeness and showing consistency. 

We thus carry out the procedure, with time-varying change of frame, recalled at the end of Appendix \ref{app_injection_locking}. From \eqref{app:eq:for3rd}, we would define
$$x_2 = x_1 - \tilde{f}_1(x_1,t) \; .$$
In fact, when expanding the resulting $\tfrac{d}{dt}x_2$ up to order 3 in the timescale separation, using \eqref{eq:app:TaylExp} up to high enough order, it turns out that all the terms involving $\tilde{f}_1$ are either oscillating or higher order than 3. We thereby obtain:
\begin{eqnarray}\label{eq:app:3rdRWA:general}
    \dot{x}_2 &=& \bar{f}(x_2) - \overline{\nabla \tilde{f}_{x_2} \cdot \tfrac{\partial}{\partial t} \tilde{f}(x_2,t)} \\
    && \nonumber + \underbrace{\tfrac{1}{2} \overline{\tilde{f}(x_2,t) \cdot \nabla\nabla_{x_2,x_2}\bar{f}(x_2) \cdot \tilde{f}(x_2,t)}}_{\bullet 1}
     - \underbrace{\overline{\tilde{f}(x_2,t) \cdot \nabla\nabla_{x_2,x_2}\tilde{f}(x_2,t) \cdot (\bar{f}(x_2)+\tfrac{\partial}{\partial t} \tilde{f}(x_2,t)) }}_{\bullet 2}\\
    && \nonumber + \underbrace{\nabla_{x_2}\bar{f}(x_2)\cdot\left(\overline{\nabla_{x_2}\tilde{f}(x_2,t) \cdot \tilde{f}(x_2,t)} \right)}_{\bullet 3} - \underbrace{\overline{\nabla_{x_2}\tilde{f}(x_2,t)\cdot\left(\nabla_{x_2}\bar{f}(x_2) \cdot \tilde{f}(x_2,t) \right)}}_{\bullet 4} \\
    && \nonumber - \underbrace{\overline{\nabla_{x_2}\tilde{f}(x_2,t)\cdot\left(\nabla_{x_2} \tfrac{\partial}{\partial t} \tilde{f}(x_2,t) \cdot \tilde{f}(x_2,t) \right)}}_{\bullet 5} \; .
\end{eqnarray}
The first row expresses the 1st and 2nd order, the next rows express the 3rd order. Here, we have used the notation $g_1 \cdot \nabla\nabla g_2 \cdot g_3 = \sum_{j,k} (g_1)_j (g_3)_k \tfrac{\partial^2}{\partial x_j \partial x_k} g_2$, while an overline again denotes time average. The labels $\bullet 1$ to $\bullet 5$ are used in the discussion below.

The reader interested in getting a hand on this method may try to compute the corresponding terms for an injection locking circuit alone (R or RC filter). Tedious computations can be reduced by directly focusing on non-rotating terms. 

For a system of the form \eqref{eq:E2} and assuming no frequency $\omega_L$ in $B_J(t)$, we have
$$ \nabla\nabla \bar{f} = 0 \text{ thus } {\bullet 1} \equiv 0  \qquad \text{and} \qquad g_1 \cdot \nabla\nabla \tilde{f} \cdot g_3 = (g_1)_{\varphi_J} (g_3)_{\varphi_J} \tfrac{\partial^2}{\partial \varphi_J^2} \tilde{f} \; .$$

In the cat-qubit dynamics \eqref{eq:app:C5mod2}, we further have $\nabla \bar{f}$ identically 0 on oscillator $A$, diagonal $-\kappa_b/2$ on oscillator $B$, and reduced to $\tfrac{\partial \bar{f}_{n_R}}{\partial n_R}=\tfrac{\partial \bar{f}_{\varphi_J}}{\partial n_R} = -1/\tau$ for the remaining components. With this, $\bullet 4$ completely cancels with the part of $\bullet 3$ concerning the buffer oscillator. Furthermore, one computes that $\bullet 5$ vanishes and the remaining resonant components, from $\bullet 2$ and $\bullet 3$, yield the following model at 3rd order averaging:
\begin{eqnarray}
\nonumber
    \tfrac{d}{dt} \alpha & \simeq & i \Delta_A \cos(\hat{\varphi}_J)\, \alpha - 2i g_2\, e^{i \,\hat{\varphi}_J} \, \alpha^* \beta \\ \nonumber
    \tfrac{d}{dt} \beta & \simeq & i \Delta_{B,1} \cos(\hat{\varphi_J})\, \beta + i \Delta_{B,2} \, \beta - \frac{\kappa_B}{2}\, \beta - i \epsilon_d - (\tfrac{\kappa_b}{4\omega_b})\, \epsilon_d - i g_2^* e^{-i \, \hat{\varphi}_J}\, \alpha^2 \\ \label{eq:app:3rdorder:shifts}
    \tfrac{d}{dt} n_R & \simeq & \frac{-n_R}{\tau} - \frac{\nu_0 \epsilon_L}{2 \tau \omega_L} \cos(\hat{\varphi}_J) - \frac{\nu_0 \epsilon_L \sin(\hat{\varphi}_J)}{2} \Big( 1 - (\delta\omega - \tfrac{n_R}{\tau})\tfrac{1}{\omega_L}\Big) \\ \nonumber
    \tfrac{d}{dt} \hat{\varphi}_J & \simeq & \frac{-n_R}{\tau} - \frac{\nu_0 \epsilon_L}{2 \tau \omega_L} \cos(\hat{\varphi}_J) + \delta\omega \; ,
\end{eqnarray}
with $\Delta_A = \frac{E_J \pza^2 \epsilon_L}{2\hbar}$, $\Delta_{B,1} = \frac{E_J \pzb^2 \epsilon_L}{2\hbar}$, $\Delta_{B,2}=(\frac{\kappa_B}{2})(\frac{\kappa_B}{4\omega_B})$ and $g_2 = \frac{E_J \pza^2 \pzb}{4 i\hbar}$ as in the main text. 

At this stage, the RC locking dynamics still appears independent of the cat stabilization circuit. Its steady state requires $\frac{-n_R}{\tau} - \frac{\nu_0 \epsilon_L}{2 \tau \omega_L} \cos(\hat{\varphi}_J)=-\delta\omega$ in the last equation, and plugging this into $\tfrac{d}{dt} n_R$ yields
$$\delta\omega = \frac{-\nu_0\epsilon_L \sin(\hat{\varphi}_J)}{2}\;\Big(1 - \tfrac{1}{\omega_L\tau} \frac{\nu_0 \epsilon_L}{2\omega_L} \cos(\hat{\varphi}_J)\Big) \; .$$
The last term inside the bracket is only a small correction, so in good approximation, locking is possible under the usual condition $|\delta\omega| < |\tfrac{\nu_0 \epsilon_L}{2}|$. 

The dynamics of the harmonic oscillators $\alpha$ and $\beta$ features essentially the same terms as \eqref{eq:eqA5}, plus some detunings. Among these, $\Delta_{B,2}$ reflects the standard damping-induced detuning on the buffer, which can be calibrated out. The two other detunings, with $\Delta_A$ and $\Delta_{B,1}$, depend on $\hat{\varphi}_J$ and hence on the DC frequency noise $\delta\omega$. Their nominal value, at $\delta\omega=0$, can be calibrated out. Their impact is also reduced by taking a smaller locking tone, e.g.~increasing $\nu_0$ and decreasing $\epsilon_L$ to keep $|\nu_0 \epsilon_L|$ constant. Even in presence of these detunings, the steady states of the harmonic oscillators, conditioned on the value of $\hat{\varphi}_J$, are of the same type as with \eqref{eq:eqA5}:
\begin{itemize}
    \item One steady state with $\alpha_{ss}=0$ and $\beta_{ss}=\frac{-2i\epsilon_d-\kappa_b\epsilon_d/(2\omega_b)}{\kappa_B - 2 i \Delta_B}$, should be unstable, with $\Delta_B = \Delta_{B,1} \cos(\varphi_J)+\Delta_{B,2}$.
    \item The other equilibria should satisfy, writing $\alpha_{ss} = r_{\alpha}\, e^{i\theta_a}$:
    $$ \Big( r_\alpha^2 + (\kappa_B - 2i\Delta_B) \tfrac{\Delta_A \cos(\hat{\varphi}_J)}{4i |g_2|^2} \Big) \, e^{2 i \theta_a} \; = \; \frac{-\epsilon_d\, e^{i \hat{\varphi}_J}}{g_2^*} \; (1-\tfrac{i\kappa_B}{4\omega_B}) \quad \text{ and } \quad \beta_{ss} = \frac{\Delta_A \cos(\hat{\varphi}_J)}{2 g_2}\, e^{2i\theta_a-i\hat{\varphi}_J} \; .$$
    Provided the detunings are not too big, this defines a pair of solutions with the same value of $\beta_{ss}$ but opposite values of $\alpha_{ss}$, thus compatible with a cat qubit. 
    
    Note how the phase $\hat{\varphi}_J(0)$ now explicitly appears in the value of the cat angle: $g_2$ now appears as the combination $g_2 e^{i \hat{\varphi}_J}$. In particular, for locking at $\hat{\varphi}_J=\pi$, the sign of $g_2$ is reversed compared to Appendix \ref{appendix:classical_EOM} and the cat angle, determined by the left steady state equation, changes by $\pi/2$. When $\hat{\varphi}_J$ deviates from this value due to noise, the cat angle follows this deviation, halved.
    
    As a second effect, if detunings deviate too much from nominal, then the cat size $r_{\alpha}^2$ also gets affected, possibly up to the point where this solution vanishes. For instance, the left equation has no solution if $\frac{|\Delta_A \cos(\hat{\varphi}_J)| \kappa_B}{4 |g_2|^2} > |\frac{\epsilon_{d}}{g_2}|$. This is a second reason to keep $|\sin(\hat{\varphi}_J)|$ small.
\end{itemize}

The effect of the cat qubit on the locking dynamics is not visible at this order, where we have assumed that all ``small'' terms were of equal importance $\varepsilon \ll 1$ and we went up to order $\varepsilon^3$ included. Like announced, this Section was meant more to show how the systematic averaging method indeed retrieves the dynamics as expected, with precise numbers, rather than to draw new conclusions.

\section{Effect of photon number \texorpdfstring{$|\alpha|^2$}{Alpha} on locking}\label{appendix:photon_number_effect}

\as{In the previous two sections, we demonstrated that averaging up to second order shows the locking mechanism, while averaging up to 3rd order makes the cat-stabilizing dynamics appear separately. In this section, we aim to explore the interaction between the cat-stabilizing and locking mechanisms. For this, instead of going yet one order higher in the general RWA, we consider our particular system and working point, and perform a change of frame to push the original dynamics to order $(\varphi_{\text{zpf}})^2$. With this, we obtain higher-order corrections after fewer computations.}


Employing the RC locking circuit model as detailed in Appendix \ref{app_injection_locking}, we write:
\begin{equation}
\begin{cases}
    \begin{aligned}
        \dot{\alpha} &= -i(\omega_a\as{+c_a}) \alpha  \ta{+} i\frac{E_J}{\hbar}\pza \sin\left(\varphi_J+\omega_L t \right)\\ 
        \dot{\beta} &= -i(\omega_b\as{+c_b}) \beta  \ta{+} i\frac{E_J}{\hbar} \pzb \sin\left(\varphi_J+\omega_L t \right) - i{\kappa_b} \text{Im}(\beta)- 2i \text{Re}{(\epsilon_d e^{-i\omega_d t})}\\
        \dot{n}_R &= -n_R/\tau + \nu_0 \sin({\varphi}_J+\omega_L t)\\
        \dot{\varphi}_J &= \delta\omega  -n_R/\tau  + \epsilon_L \omega_L \cos(\omega_L t)  -  2\pza \omega_a \text{Im}(\alpha) - 2\pzb \omega_b \text{Im}(\beta) \label{eq:eqA1c} \; .
\end{aligned}
\end{cases}
\end{equation}
\as{Here $c_a$ and $c_b$ are frequency shifts meant to counter the $O(\varphi_{\text{zpf}})^3$ induced detunings observed in Eq.~\eqref{eq:app:3rdorder:shifts}. For a known $\delta\omega$ and thus $\hat{\varphi}_J$ in steady state, they can be precomputed. In practice, they will be automatically included when calibrating the experiment's best possible behavior (provided $\delta\omega$ remains constant over the calibration time).}\\

\as{We focus on timescale separations governed by $\epsilon << 1$, assuming all other prefactors to be less significant. As already mentioned, this viewpoint is quite arguable for the values taken in our simulations and envisioned in experiments, since we will increase $\epsilon$ as much as possible to better protect the cat-qubit. However, it provides a valid analysis for some parameter settings at least, which can serve as a basis for further investigation. Explicitly, we will assume the following.
\begin{itemize}
    \item $E_J/\hbar$, $\omega_a$, $\omega_b$, $\omega_L$ are all of the same order and $\alpha,\beta$ are at most of \underline{order 1}; $\nu_0/\omega_a$, $\pza$, $\pzb$ and $\epsilon_L$ are all of \underline{order $O(\epsilon)$};  $1/\tau$, $\delta\omega$ (see the locking condition at first order) and $\kappa_b$ (to maximize cat-qubit protection, which scales as $1/\kappa_b$ when $\kappa_b \gg |g_2| = O(\epsilon^3)$) are all of \underline{order $O(\epsilon^2)$} ;  while $\epsilon_d$ (see how it governs $|\alpha_{ss}|$ at first order) is of \underline{order $O(\epsilon^3)$} and later $c_a$ and $c_b$ will also be chosen of that order at most.
    \item To perform a Taylor expansion of the sine nonlinearity, we assume that, in stationary regime, $\varphi_J = \varphi_{ss}+\delta\varphi_{J,1}(t)$ with $\varphi_{ss}$ constant and $\delta\varphi_{J,1}(t)=O(\epsilon)$.
    \item Next, towards making the right-hand side of Eq.\eqref{eq:eqA1c} of order $O(\epsilon^2)$, we perform displacements like in the previous sections, defining $\alpha = \alpha_{\text{new}}+\xi_a$ and similarly for $\beta,n_R$. By choosing
\begin{align*}
    \xi_{a,b} &= \frac{-\varphi_{\text{zpf}; a,b} E_J/2\hbar}{i(\omega_{a,b}+c_{a,b}-\omega_L)}\, e^{-i(\omega_{L}t+\varphi_{ss})} + \frac{\varphi_{\text{zpf}; a,b} E_J/2\hbar}{i(\omega_{a,b}+c_{a,b}+\omega_L)}\, e^{i(\omega_{L}t+\varphi_{ss})} \; ,\\
    \xi_R &= \frac{-\nu_0/2i}{1/\tau - i \omega_L} \, e^{-i(\omega_{L}t+\varphi_{ss})} + \frac{\nu_0/2i}{1/\tau + i \omega_L} \, e^{i(\omega_{L}t\varphi_{ss}) \; ,}
\end{align*}
which are of order $O(\epsilon)$, we cancel the order-0 term in the Taylor expansion of $\sin\left(\varphi_{ss}+\omega_L t + \delta\varphi_{J,1}\right)$ with respect to $\delta\varphi_{J,1}$.
\item After these displacements, we go to a rotating frame at $\omega_a$ and $\omega_b$ respectively for $\alpha_{\text{new}}$ and $ \beta_{\text{new}}$.
\item In this rotating frame, we assume a decomposition into steady state value and $O(\epsilon)$ variation, like we have done for $\varphi_J$, now for $\alpha_{\text{new}}, \beta_{\text{new}}$ and $n_{R,\text{new}}$. While the steady state values $\alpha_{ss}$ and $\varphi_{ss}$ will be fixed later, we already assume here that $\beta_{ss}=0$ (a key property for our target of cat-qubit stabilization) and that $n_{R,ss}/\tau = \delta\omega$ (as observed in the lower order locking condition). Deviations from the latter values will appear as biases in $\delta\beta$ and $\delta n_R$; we will check a posteriori that those indeed remain small, so our assumption was consistent.
\item Plugging all these into $\delta\dot{\varphi}_{J,1}$, we obtain a right-hand side with several oscillating terms which do not depend on the variables. To get rid of those terms, we define $\delta\varphi_{J,1} = \delta\varphi_J + \xi_J$ where
\begin{eqnarray*}
    \xi_J &=& \epsilon_L \sin(\omega_L t) - \pza (\alpha_{ss} e^{-i\omega_a t} + \alpha_{ss}^* e^{i\omega_a t}) + s_J \sin(\omega_L t + \varphi_{ss}) + c_J \cos(\omega_L t + \varphi_{ss}) \quad \text{where} \\
    && s_J \;=\; \tfrac{\omega_a \pza^2 E_J}{\hbar \omega_L} \left( \tfrac{1}{\omega_a+c_a+\omega_L}-\tfrac{1}{\omega_a+c_a-\omega_L} \right) +\tfrac{\omega_b \pzb^2 E_J}{\hbar \omega_L} \left( \tfrac{1}{\omega_b+c_b+\omega_L}-\tfrac{1}{\omega_b+c_b-\omega_L} \right) + \tfrac{\nu_0}{\omega_L^2 \tau(1+1/(\omega_L\tau)^2)} \;\; , \\
    && c_J \;=\; \tfrac{\nu_0}{\omega_L^3 \tau^2} \; . 
\end{eqnarray*}
Keep in mind for later that $s_J = O(\epsilon^2)$ and $c_J = O(\epsilon^5)$. We then get the following starting point for our analysis:
\begin{eqnarray}
    \delta\dot{\alpha} &=& -i c_a (\delta\alpha+\alpha_{ss}) + \tfrac{i E_J \pza}{\hbar} e^{i\omega_a t} \Big( \sin(\omega_L t + \varphi_{ss}) \cos_1(\delta\varphi_J+\xi_J)+\cos(\omega_L t + \varphi_{ss}) \sin(\delta\varphi_J+\xi_J)\Big) \nonumber\\
    \delta\dot{\beta} &=& -i c_b \delta\beta + \tfrac{i E_J \pzb}{\hbar} e^{i\omega_b t} \Big( \sin(\omega_L t + \varphi_{ss}) \cos_1(\delta\varphi_J+\xi_J)+\cos(\omega_L t + \varphi_{ss}) \sin(\delta\varphi_J+\xi_J)\Big) \nonumber\\
    && \phantom{KKKK} - i \epsilon_d (1+e^{2i\omega_b t}) - \tfrac{\kappa_b}{2}(\delta\beta - \delta\beta^* + \xi_b e^{i\omega_b t} - \xi_b^* e^{i\omega_b t}) \nonumber \\
    \delta\dot{n}_R &=& -\delta n_R / \tau - \delta\omega + \nu_0 \Big( \sin(\omega_L t + \varphi_{ss}) \cos_1(\delta\varphi_J+\xi_J)+\cos(\omega_L t + \varphi_{ss}) \sin(\delta\varphi_J+\xi_J)\Big) \label{eq:appG:afterallframes} \\
    \delta\dot{\varphi}_J &=& -\delta n_R / \tau - 2 \pza \omega_a \text{Im}(\delta\alpha\, e^{-i\omega_a t}) - 2 \pzb \omega_b \text{Im}(\delta\beta\, e^{-i\omega_b t}) \; , \nonumber
\end{eqnarray}
where $\cos_1(x) = \cos(x)-1$.
\item We will now perform a Rotating Wave Approximation on Eq.\eqref{eq:appG:afterallframes}, known in classical dynamics as ``averaging'', along the same lines as in Appendix \ref{app_injection_locking}. This amounts to performing a change of frame involving the small oscillating vector field in the right-hand side of Eq.\eqref{eq:appG:afterallframes}. Now consider this: 
\begin{itemize}
\item We will assume that all the variables $(\delta\alpha,\delta\beta,\delta n_R, \delta\varphi_J)$ are of order $O(\epsilon^2)$, and check it a posteriori. The aim is to have the right-hand side of Eq.\eqref{eq:appG:afterallframes} of order $O(\epsilon^2)$; although some of its derivatives with respect to the variables will be of order  $O(\epsilon)$.
\item Under this assumption, the change of frame \eqref{eq:app:AvCF1} from $x=(\delta\alpha,\delta\beta,\delta n_R, \delta\varphi_J)$ to $x_1$, in order to perform averaging, will be $O(\epsilon^2)$ close to identity. When iterating the procedure, the changes of frame will be even closer to identity. Hence, proving that the original variables remain of order $O(\epsilon^2)$ amounts to proving that the new variables remain of order $O(\epsilon^2)$.
\item After two changes of frame of the form \eqref{eq:app:AvCF1}, assuming variables are of order $O(\epsilon^2)$, we will obtain a system where the RWA part (constant vector field) exponentially converges to $0$ with eigenvalues of order $\epsilon^3$ (consistent with Appendix \ref{App:3rdorder}); while the oscillating vector field has been pushed to order $O(\epsilon^6)$. This implies that indeed the variables of the true system, in this frame, remain smaller than $O(\epsilon^2)$ (see e.g. \cite{sanders2007averaging} for the proof technique). By the previous point, the original variables $x=(\delta\alpha,\delta\beta,\delta n_R, \delta\varphi_J)$ thus also remain $O(\epsilon^2)$ at most, justifying our initial assumption.
\end{itemize}
\item We next perform higher-order RWA as explained in Appendix \ref{ref:AppERWA}, with two orders of change of frame. Taking into account the orders of vector fields and of their gradients, Taylor expansions like \eqref{eq:app:3rdRWA:general} are carried out up to order $O(\epsilon^5)$ terms included, keeping non-oscillating terms only, without too much difficulty. We obtain the model:
\begin{eqnarray} \nonumber
    \delta\dot{\alpha} &=& - \tfrac{E_J\pza^2\pzb }{2\hbar}\alpha_{ss}^* e^{i (\varphi_{ss}+\delta\varphi_J)} \delta\beta\\
    \nonumber && - i c_a \, (\alpha_{ss} +\delta\alpha) + \tfrac{i E_J \pza^2}{2 \hbar} \big(s_J+\cos(\varphi_{ss}+\delta\varphi_J)\epsilon_L\big)\, ( \alpha_{ss} + \delta\alpha) \\
    \nonumber && - \Big( \tfrac{i E_J \pza^4 |\alpha_{ss}|^2\epsilon_L}{4\hbar} + \tfrac{i E_J \pza^2\epsilon_L^3}{16\hbar} \Big) \cos(\varphi_{ss}+\delta\varphi_J) \, (\alpha_{ss} +\delta\alpha) \\
    \nonumber && - \Big[ \tfrac{iE_J \nu_0 \pza^2}{4 \hbar \tau} \left( \tfrac{1}{(\omega_L-\omega_a)^2} + \tfrac{1}{(\omega_L+\omega_a)^2}\right) \\
    \nonumber && \phantom{-}+ \left( \tfrac{E_J}{2\hbar}\right)^2 i\pza^4 \omega_a \left( \tfrac{1}{(\omega_L+\omega_a)\omega_L} + \tfrac{1}{(\omega_L-\omega_a)\omega_L}+ \tfrac{1}{(\omega_L+\omega_a)(\omega_L+2\omega_a)} + \tfrac{1}{(\omega_L-\omega_a)(\omega_L-2\omega_a)} \right) \\
    \nonumber && \phantom{-}+ \left( \tfrac{E_J}{2\hbar}\right)^2 i\pza^2 \pzb^2 \omega_b \left( \tfrac{1}{(\omega_L+\omega_a)(\omega_L+\omega_a-\omega_b)} + \tfrac{1}{(\omega_L-\omega_a)(\omega_L+\omega_b-\omega_a)}+ \tfrac{1}{(\omega_L+\omega_a)(\omega_L+\omega_a+\omega_b)} + \tfrac{1}{(\omega_L-\omega_a)(\omega_L-\omega_a-\omega_b)} \right) \Big]\\[2mm]
    \nonumber && \phantom{KKK} \cdot (\alpha_{ss} +\delta\alpha) \\[4mm]
    \nonumber \delta\dot{\beta} &=& \tfrac{E_J \pza^2\pzb }{2\hbar} \alpha_{ss} e^{-i(\varphi_{ss}+\delta\varphi_J)} \delta\alpha\\
    \label{eq:appGbigRWA} && -i c_b \delta \beta + \tfrac{i E_J \pzb^2 \epsilon_L e^{i (\varphi_{ss}+\delta\varphi_J)}}{2\hbar}\, \delta\beta  - \tfrac{\kappa_b}{2} \delta\beta \\
    \nonumber && - i \epsilon_d- \tfrac{\kappa_b \epsilon_d}{4\omega_b} - \tfrac{i\kappa_b E_J \pza^2 \pzb \alpha_{ss}^2 e^{-i (\varphi_{ss}+\delta\varphi_J)}}{16 \omega_b \hbar} \\
    \nonumber && + \tfrac{E_J \pza^2\pzb \alpha_{ss}^2 e^{-i (\varphi_{ss}+\delta\varphi_J)}}{4 \hbar} - \tfrac{E_J \pza^2\pzb \alpha_{ss}^2}{4\hbar}\big(\tfrac{\pza^2|\alpha_{ss}|^2 e^{-i(\varphi_{ss}+\delta\varphi_J)}}{3} + \tfrac{\epsilon_L^2}{2}\cos(\varphi_{ss}+\delta\varphi_J) \big) \\[4mm]
   \nonumber \delta \dot{n}_R &=& \tfrac{-1}{\tau} \delta n_R - \delta\omega - \tfrac{\nu_0 \epsilon_L}{2} \sin(\varphi_{ss}+\delta\varphi_J) \, \left(1-\pza^2|\alpha_{ss}|^2 - \tfrac{\epsilon_L^2}{8} + \tfrac{\epsilon_L E_J \cos(\varphi_{ss}+\delta\varphi_J)}{\hbar}(\tfrac{\pza^2}{\omega_a}+\tfrac{\pzb^2}{\omega_b}) \right) \\[4mm]
   \nonumber \delta\dot{\varphi}_J &=& \tfrac{-1}{\tau} \delta n_R - \tfrac{\pzb^2 \kappa_b E_J \epsilon_L}{2\hbar \omega_b} \sin(\varphi_{ss}+\delta\varphi_J) \\[4mm]
   \nonumber && + O(\epsilon^6) \;\; ,
\end{eqnarray}
where we have kept some higher-order terms to highlight preserved groupings.
\end{itemize}

From the model \eqref{eq:appGbigRWA}, we can carry out the following analysis for the steady state.
\begin{itemize}
    \item Requesting $\delta\dot{\varphi}_J=0$ fixes the value of $\tfrac{-1}{\tau} \delta n_R$, taking $\kappa_b$ and $1/\tau$ of the same order we have  $\delta n_{R,ss} = O(\epsilon^3)$.
    \item Then requesting $\delta \dot{n}_R = 0$ gives an equation of the type $\delta \omega = M \sin(\varphi_{ss}+\delta\varphi_J)$. The coefficient $M$ has a dominant component $\tfrac{\nu_0 \epsilon_L}{2} = O(\epsilon^2)$, but it is reduced at order $O(\epsilon^4)$ amongst others by $\tfrac{\nu_0 \epsilon_L}{2} \pza^2 |\alpha_{ss}|^2$. This is the value reported on Fig.\ref{fig:fig3} in the main text, when we bound how big a $\delta\omega$ we can reject.
    \item In the equation for $\delta\dot{\beta}$, we can calibrate $c_b,\kappa_b$ to cancel term between them on the second line. Imposing a target value of $|\alpha_{ss}|$, with $\varphi_{ss}$ fixed from the previous point, we can adjust $\epsilon_d$ to have the same amplitude as all the other constant terms in $\delta\dot{\beta}$. The angle $\theta_a$ of $\alpha_{ss}$ adjusts itself such that all the constant tems cancel.
    \item In the equation for $\delta\dot{\alpha}$, besides the first line, all terms are multiples of $(\alpha_{ss}+\delta\alpha)$. With the rest fixed, we can calibrate $c_a$ in order to cancel them all.
    \item In conclusion, we get a steady state with $\delta\alpha=\delta\beta=\delta\varphi_{ss} = 0$ and $\delta n_{R,ss} = O(\epsilon^3)$, provided essentially we have calibrated well and $\frac{\nu_0 \epsilon_L J^*}{2}  > \delta\omega$ with $J^* = 1-\frac{\epsilon^2}{8}-\pza^2 |\alpha_{ss}|^2$.
\end{itemize}
The first line of $\delta\dot{\alpha}$ and of $\delta\dot{\beta}$ are the target cat-stabilizing dynamics.

Finally, we can examine the stability of this model. 
\begin{itemize}
\item It is block-diagonal, with the cat-dynamics $(\delta\alpha,\delta\beta)$ having their own eigenvalues and the junction stabilization $(\delta n_R,\delta\varphi_J)$ their own ones.
\item The cat-qubit part has the same eigenvalues as without locking, irrespectively of the value of $\varphi_{ss}$ (which makes physical sense), provided the same cat amplitude $\alpha_{ss}$ is calibrated.
\item The locking dynamics features the dominant behavior as highlighted in Appendix \ref{app_injection_locking}, with eigenvalues $\lambda = \frac{-1}{2\tau} \pm \sqrt{(\tfrac{1}{2\tau})^2 + \frac{\nu_0 \epsilon_L J^* \cos(\varphi_{ss})}{2 \tau}}$ being stable only if $\cos(\varphi_{ss}) < 0$. In particular, the locking rate becomes slower and slower as $\cos(\varphi_{ss})$ approaches $0$ to reach bigger $|\sin(\varphi_{ss})|$ for countering bigger $\delta\omega$.
\end{itemize}
}

\section{Derivation of the Kerr terms}\label{appendix:Kerr}

In this section, we analytically quantify the Kerr terms in order to compare their strength with that of the two-photon exchange $g_2$.\\

\as{\paragraph*{Methodology:} In this paper, we focus on deriving approximate dynamics at various orders of $\varphi_{\text{zpf},a/b}$, although this is of course not the only parameter. Our $g_2$ is already of order $\varphi_{\text{zpf},a/b}^3$. In order to trust conclusions up to order $\varphi_{\text{zpf},a/b}^5$ included, we must do two things compared to Appendix \ref{appendix:eff_Hamiltonian}. 
\begin{itemize}
    \item The first point is of course to carry out the Taylor expansions up to order $\varphi_{\text{zpf},a/b}^5$ included.
    \item The second point is to carry out the RWA to second order. Indeed, the first-order RWA, consisting in neglecting all oscillating terms of $H$ in the rotating frame, is only accurate up to errors of order $(\Vert H \Vert / \omega_a)^2 \simeq \varphi_{\text{zpf},a/b}^4$. By performing a second-order RWA, the error is pushed to order $\varphi_{\text{zpf},a/b}^6$. The formula for the effective Hamiltonian at 2nd order RWA is: 
\begin{equation}\label{eq:RWA2:Pierreformula}
        H_{\text{eff},2} = H_{\text{eff},1} - i \overline{(H(t)-H_{\text{eff},1})\left(\int_t(H(t)-H_{\text{eff},1}) dt \right)}\; ,
    \end{equation}
    where $H_{\text{eff},1}$ denotes the averaged Hamiltonian at first order (``usual'' RWA), the integral denotes the primitive of the oscillating Hamiltonian over time (not to be confused with the much-harder-to-compute propagator associated to this Hamiltonian), and overline denotes keeping the non-oscillating terms only \cite{RMcourse}.

    Note that for the second term in Eq.\eqref{eq:RWA2:Pierreformula}, i.e.~the  2nd order RWA correction, it is sufficient to consider the Taylor expansion of $H$ up to order $\varphi_{\text{zpf},a/b}^3$.
\end{itemize}}

\as{\paragraph*{Results:}} We first consider the case where the DC-voltage source is perfect \as{and there is no injection locking circuit.
The starting point is the higher-order Taylor version of Eq.\eqref{eq:A7:Hrot}, namely the Hamiltonian} 
\begin{align}\label{eq:eqH6}
    H = & 
    - E_J \cos ( \omega_{dc}t) \cos \left( \pza ( a^\dag e^{i \omega_a t} + a e^{-i \omega_a t} + \xi_a^{*(t)} + \xi_a^{(t)} ) + \vphantom{a^\dagger} \pzb ( b^\dag e^{i \omega_b t} + b e^{-i \omega_b t} + \xi_b^{*(t)} + \xi_b^{(t)} ) \right)  \nonumber \\
    & - E_J \sin ( \omega_{dc}t) 
 \ta{\sin_1}\left( \pza ( a^\dag e^{i \omega_a t} + a e^{-i \omega_a t} + \xi_a^{*(t)} + \xi_a^{(t)} )  
     + \vphantom{a^\dagger} \pzb ( b^\dag e^{i \omega_b t} + b e^{-i \omega_b t} + \xi_b^{*(t)} + \xi_b^{(t)} ) \right)  \nonumber \\
     & \as{+~2\text{Re}\left(\hbar 
    \epsilon_d
    e^{-i\omega_d t}\right)(b^\dag e^{i \omega_b t} + b e^{-i \omega_b t} ) \; ,}
\end{align}
\ta{where $\sin_1$ denotes the sine function with the first-order term removed i.e: $\sin_1(x) = \sin(x) - x$}. The nonlinear terms can be expanded in normal order like in Appendix \ref{appendix:eff_Hamiltonian}, keeping terms up to 5th order included in $\pza, \pzb$ and $\xi_a, \xi_b$. \as{Then the non-rotating terms are
\begin{align} 
H_{\text{eff,}1}/\hbar &= {\delta_a} a^{\dagger} a + {\delta_b} b^{\dagger} b  + \left(g_2 a^{\dagger 2} b + \ta{g_2^a  a^{\dagger 2}(a^{\dagger}a) b + g_2^b (b^{\dagger}b) a^{\dagger 2} b }+ \epsilon_d b^{\dagger} + \ta{h.c} \right)\nonumber + O(\varphi_{\text{zpf,}a/b})^6 \; ,  \\[2mm] 
 \text{with } \;\;\; \delta_{a,b} &= \frac{\tilde{E}_J}{\hbar}\varphi_{\text{zpf},a,b}^2 \left( \pza\text{Im}(\bar{\xi}_a) + \pzb \text{Im}(\bar{\xi}_b) \right) \; ; \;\;
\ta{g_2^a = \frac{-\tilde{E}_J}{12i\hbar} \pza^4 \pzb \; ; \;\;    g_2^b = \frac{-\tilde{E}_J}{8i\hbar} \pza^2 \pzb^3.} \nonumber
\end{align}
See Appendix \ref{appendix:eff_Hamiltonian} for the definition of $\bar{\xi}_{a,b}$. These are frequency renormalizations of order $(\varphi_{\text{zpf,}a/b})^4$ and the terms $g_2^a, g_2^b$ described in Appendix \ref{appendix:eff_Hamiltonian}. 
In the second part of Eq.\eqref{eq:RWA2:Pierreformula}, at this order the only remaining non-rotating terms also induce frequency renormalizations, namely:
\begin{align}\nonumber
    \frac{H_{\text{eff},2}}{\hbar} &= \frac{H_{\text{eff},1}}{\hbar} + \delta_{a,2}\; a^\dagger a + \delta_{b,2}\; b^\dagger b \\ \nonumber
        & \text{with }\;\; \delta_{a,2} = 
    \left(\tfrac{\tilde E_J}{2\hbar}\right)^2 \pza^2\left( \tfrac{\pza^2}{\omega_b- 4\omega_a} - \tfrac{\pza^2}{\omega_b} - \tfrac{\pzb^2}{2\omega_b - \omega_a} - \tfrac{\pzb^2}{2\omega_b - 3\omega_a} - \tfrac{4\pzb^2}{3\omega_a} \right) \; , \\
    & \phantom{\text{with }} \;\; \delta_{b,2} = \left(\tfrac{\tilde E_J}{2\hbar}\right)^2 \pzb^2\left( 
    \tfrac{-\pzb^2}{\omega_b+2\omega_a} - \tfrac{\pzb^2}{3\omega_b-2\omega_a}- \tfrac{\pza^2}{2\omega_b - \omega_a} + \tfrac{\pza^2}{2\omega_b - 3\omega_a} + \tfrac{2\pza^2}{3\omega_a}  \right)
 \; .
\end{align}}

\as{As long as there are no quasi-resonances at this order, other effects like Kerr and cross-Kerr can thus only appear at higher orders, meaning at least $O(\varphi_{\text{zpf}})^6$.\\}

Let us now \as{add the locking tone, assuming that} the Josephson junction phase is already locked \as{and} has the following form:
\begin{equation*}
\varphi_J(t) = \omega_{dc}t + \phi_{0} + c_J \cos(\omega_{dc}t + \phi_{0} ) , 
\end{equation*}
where $c_J$, $\phi_{0}$, depend on the drives $\epsilon_d(t)$ and $\epsilon_L(t)$, but $c_J \ll 1$. \as{The predominance of frequency $\omega_{dc}$ has indeed been observed in simulations.} For the sake of simplicity and without loss of generality, let us put $\phi_0 = 0$.
We can now rewrite the Hamiltonian as : 
\begin{align}\label{eq:eqH13}
    H = & \hbar\omega_a a^{\dagger} a + \hbar \omega_b b^{\dagger} b 
\as{ \; +~2\text{Re}\left(\hbar 
    \epsilon_d
    e^{-i\omega_d t}\right)(b^\dag + b) \; ,}  \\    
    &- E_J \cos \left[ \vphantom{a^{\dagger}} \omega_{dc}t + c_J \cos ( \omega_{dc}t ) \right. \nonumber - \pza ( a^{\dagger} + a + \bar{\xi}_a^* e^{i\omega_{dc} t } + \bar{\xi}_a e^{-i\omega_{dc} t} )  \nonumber  - \left. \vphantom{a^\dagger} \pzb ( b^{\dagger} + b + \bar{\xi}_b^* e^{i\omega_{dc} t } + \bar{\xi}_b e^{-i\omega_{dc} t} ) \vphantom{a^{\dagger}} \right] 
\end{align}
\as{where $\bar{\xi}_a = \xi_{a1}+\xi_{a2}^*$, and similarly for $\bar{\xi}_b$. Indeed, considering $c_J$ of order $\varphi_{\text{zpf}}$ at most, the Taylor first order is unchanged up to a scalar term, yielding the same $\xi_a^{(t)}$ and $\xi_b^{(t)}$ as previously. Moreover, $c_J$ appears in \eqref{eq:eqH13} just as a correction to the displacement amplitude:
\begin{align}
 \pza \bar{\xi}_a + \pzb \bar{\xi}_b \;\; \text{ becomes } \;\; \pza \bar{\xi}_a + \pzb \bar{\xi}_b + c_J/2 \; .
\end{align}
Performing the same calculations as without the locking signal, we have two cases.
\begin{itemize}
\item If $c_J$ is of order $(\varphi_{\text{zpf}})^2$, like $\pza \bar{\xi}_a + \pzb \bar{\xi}_b$, then the qualitative picture does not change. Only the values of frequency renormalization and of $g_2^a$, $g_2^b$ possibly change at this order.
\item If $c_J$ is considered larger than $(\varphi_{\text{zpf}})^2$, then some terms previously discarded may have to be kept. One checks that no new terms appear from 2nd order RWA, i.e.~in the second part of Eq.\eqref{eq:RWA2:Pierreformula}.
\end{itemize}
In both cases, there are thus no Kerr terms up to order $O(\varphi_{\text{zpf}})^6$. The values are adjusted as follows.
}

\ta{
In first-order RWA: 

\begin{align}
    \delta_{a,b} &= \frac{\tilde{E}_J}{\hbar} \varphi_{\text{zpf,a,b}}^2(J_0(c_J) + J_1(c_J))(\pza \text{Im}(\bar{\xi}_a) + \pza\bar{\xi}_b)\\
    g_2 &= \frac{E_J}{4i\hbar}(J_0(c_J) + J_1(c_J))\pza^2\pzb \\
    g_2^{a} &= -\frac{\tilde{E}_J}{12i\hbar}(J_0(c_J) + J_1(c_J))\pza^4\pzb \\
    g_2^{b} &= -\frac{\tilde{E}_J}{8i\hbar}(J_0(c_J) + J_1(c_J))\pza^2\pzb^3
\end{align}

In second-order RWA: 
\begin{align}
    \delta_{a,2} &=  \left(\tfrac{\tilde E_J}{2\hbar}\right)^2 J_0(c_J)(J_0(c_J) -2 J_1(c_J))\pza^2\left( \tfrac{\pza^2}{\omega_b- 4\omega_a} - \tfrac{\pza^2}{\omega_b} - \tfrac{\pzb^2}{2\omega_b - \omega_a} - \tfrac{\pzb^2}{2\omega_b - 3\omega_a} - \tfrac{4\pzb^2}{3\omega_a} \right) \\   
    \delta_{b,2} &= \left(\tfrac{\tilde E_J}{2\hbar}\right)^2 J_0(c_J)(J_0(c_J) -2 J_1(c_J))\pzb^2\left( 
    \tfrac{-\pzb^2}{\omega_b+2\omega_a} - \tfrac{\pzb^2}{3\omega_b-2\omega_a}- \tfrac{\pza^2}{2\omega_b - \omega_a} + \tfrac{\pza^2}{2\omega_b - 3\omega_a} + \tfrac{2\pza^2}{3\omega_a}  \right)
\end{align}
}

\section{Practical limitation in using a Josephson voltage standard source}\label{appendix:standard}
\begin{figure}
    \centering
    \includegraphics[width=0.5\linewidth]{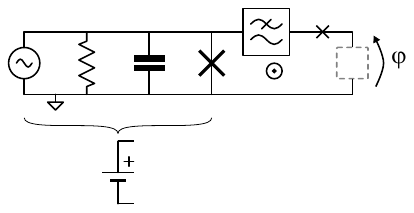}
    \caption{Lumped-element model of a junction biased using an ideal voltage source. This voltage source comprises a microwave source, a capacitor, a resistor and a junction. In the right parameter regime, a voltage proportional to the microwave source frequency develops across the junction, which can in turn be used to bias the mixing junction of our circuit (on the right). The voltage and underlying phase are locked to the microwave source, ensuring no noise at this point. However, to transfer this phase to the mixing junction, the connection needs to be superconducting, resulting in a possibly large superconducting loop. This loop may pick up a large amount of noise, decreasing the interest of this approach.}
    \label{fig:ideal_voltage}
\end{figure}

In figure~\ref{fig:ideal_voltage}, we quickly explain how a Josephson voltage standard source could be used to provide a perfect DC bias. Although this could work in principle, a flux noise is necessarily picked up which decreases the interest of this approach.

\section{Frequency parameter choice}\label{appendix:frequencies}

\begin{figure}[h]
    \centering
    \includegraphics[width=1\linewidth]{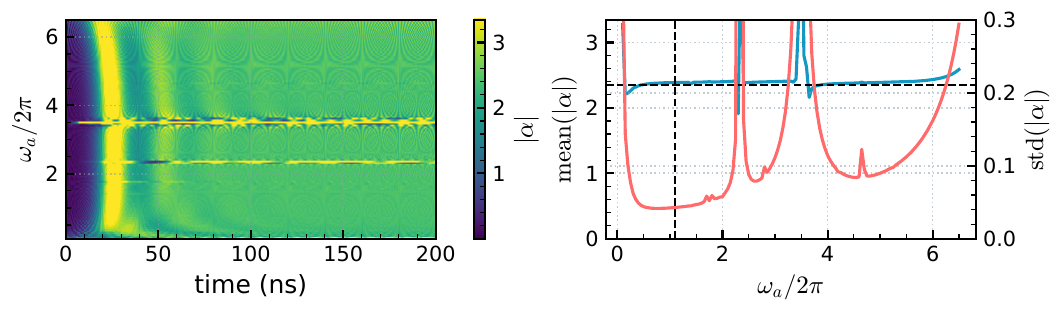}
    \caption{\textbf{Choosing $\omega_a$ (left plot)} Evolution of $|\alpha|$ (color) under the ideal dynamics as a function of time and $\omega_a$. The buffer drive amplitude is set such that the cat size for the ideal dynamics would be $\sqrt{5.5}=2.34$, corresponding to green color. The steady state is reached after a 150 ns evolution, and some spurious resonances are already apparent. \textbf{(right plot)} Mean value (blue) and standard deviation (red) of the last 20 ns of the trajectories used for the left plot, as a function of $\omega_a$. To be close to the reduced dynamics, we aim for a mean value of $|\alpha|$ close to $\sqrt{5.5}$ (dashed horizontal line) and a minimal standard deviation. Our final choice is $\omega_a/2\pi = 1.1$ GHz (dashed vertical line).}
    \label{fig:steady_alpha_std}
\end{figure}

In this paragraph, we outline the phenomenological method employed to optimize the frequency selection for $\omega_a, \omega_b$ and $\omega_p$. We initially set the memory mode to a low frequency in order to leverage the larger intrinsic lifetime, as the product $\omega_a T_1$ is typically constant experimentally. Subsequently, the pump frequency is fixed at 7 GHz, a relatively high value \as{to induce a large $\omega_b$ and hence a low-temperature buffer mode.}
In Fig.~\ref{fig:steady_alpha_std}, we sweep $\omega_a/2\pi$ between 0.1 and 6.5 GHz, adjusting $\omega_b/2\pi$ accordingly to maintain $\omega_p/2\pi=7$ GHz. \as{The whole picture is indeed invariant when multiplying all frequencies by the same constant.} We then simulate the \as{classical} system dynamics under ideal voltage source conditions (negligible resistance, no locking tone). The optimal frequency selection should achieve the desired cat-stabilizing dynamics while minimizing spurious interactions. 
We pragmatically translate these conditions to a mean value of $|\alpha|$ in the steady state close to the value predicted by the reduced dynamics, and with a small residual standard deviation around the steady state. Fig.\ref{fig:steady_alpha_std} clearly indicates resonances that should be avoided. A minimum in the standard deviation is observed between 0.5 and 1.5 GHz, indicating that this range is sufficiently far from most resonances. Ultimately, we have chosen $\omega_a/2\pi=1.1$ GHz. 

